\documentclass[prb,raggedbottom,showpacs,twocolumn,superscriptaddress,floatfix]{revtex4}

\usepackage{graphicx}
\usepackage{amssymb}
\usepackage{amsmath}
\usepackage{amscd}
\usepackage{eucal}
\usepackage{color}
\usepackage{bm}

\newcommand{\e}{{\rm e}}
\newcommand{\rmd}{{\rm d}}
\newcommand{\rmi}{{\rm i}}

\newcommand{\half}{{\textstyle{\frac{1}{2}}}}

\newcommand{\eps}{\epsilon}

\newcommand{\eminus}{ e^{\operatorname{-}} }

\newcommand{\kB}{k_{\rm B}}

\newcommand{\ignore}[1]{\relax}

\definecolor{DarkGreen}{rgb}{0,0.7,0}

\begin{document}

\title{Finding the quantum thermoelectric with maximal efficiency \\
and minimal entropy production at given power output}

\author{Robert S.~Whitney}
\affiliation{
Laboratoire de Physique et Mod\'elisation des Milieux Condens\'es (UMR 5493), 
Universit\'e Grenoble 1 and CNRS, Maison des Magist\`eres, BP 166, 38042 Grenoble, France.}

\date{March 16, 2015}
\begin{abstract}
We investigate the nonlinear scattering theory for quantum systems
with strong Seebeck and Peltier effects, and consider their use as
heat-engines and refrigerators with finite power outputs.
This article gives detailed derivations of the results summarized in Phys.\ Rev.\ Lett.\ {\bf 112}, 130601 (2014).   
It shows how to use the scattering theory to find 
(i) the quantum thermoelectric with maximum possible power output,
and (ii) the quantum thermoelectric with maximum efficiency at given power output.  
The latter corresponds to a minimal entropy production at 
that power output.
These quantities are of quantum origin since they depend on system size over electronic wavelength, 
and so have no analogue in classical thermodynamics.
The maximal efficiency coincides with Carnot efficiency at zero power output, but decreases with increasing power output. This gives a fundamental lower bound on entropy production, which means that reversibility (in the thermodynamic sense) is impossible for finite power output.
The suppression of efficiency by (nonlinear) phonon and photon effects is addressed in detail; 
when these effects are strong, maximum efficiency coincides with maximum power.
Finally, we show in particular limits (typically without magnetic fields) that relaxation within the quantum system does not allow the system to exceed the bounds derived for relaxation-free systems, however, a general proof of this remains elusive.  
\end{abstract}



\maketitle

\section{Introduction}
Thermoelectric effects in nanostructures \cite{Pekola-reviews,Casati-review,Sothmann-Sanchez-Jordan-review,Haupt-review} and molecules \cite{Paulsson-Datta2003,Reddy2007}
are of great current interest. They might enable
efficient electricity generation and refrigeration \cite{books,DiSalvo-review,Shakouri-reviews}, and could also lead to new types of sub-Kelvin refrigeration, 
cooling electrons in solid-state samples to lower temperatures than with conventional cryostats 
\cite{Pekola-reviews}, 
or cooling fermionic atomic gases \cite{Grenier2012,Brantut-Grenier-et-al2013,Grenier2014}. 
However, they are also extremely interesting at the level of fundamental physics,
since they allow one to construct the simplest possible quantum machine that converts heat flows into useful work (electrical power in this case) or vice versa.
This makes them an ideal case study for {\it quantum thermodynamics}, 
i.e. the thermodynamics of quantum systems \cite{QuantumThermodyn-book}.

\begin{figure}
\includegraphics[width=\columnwidth]{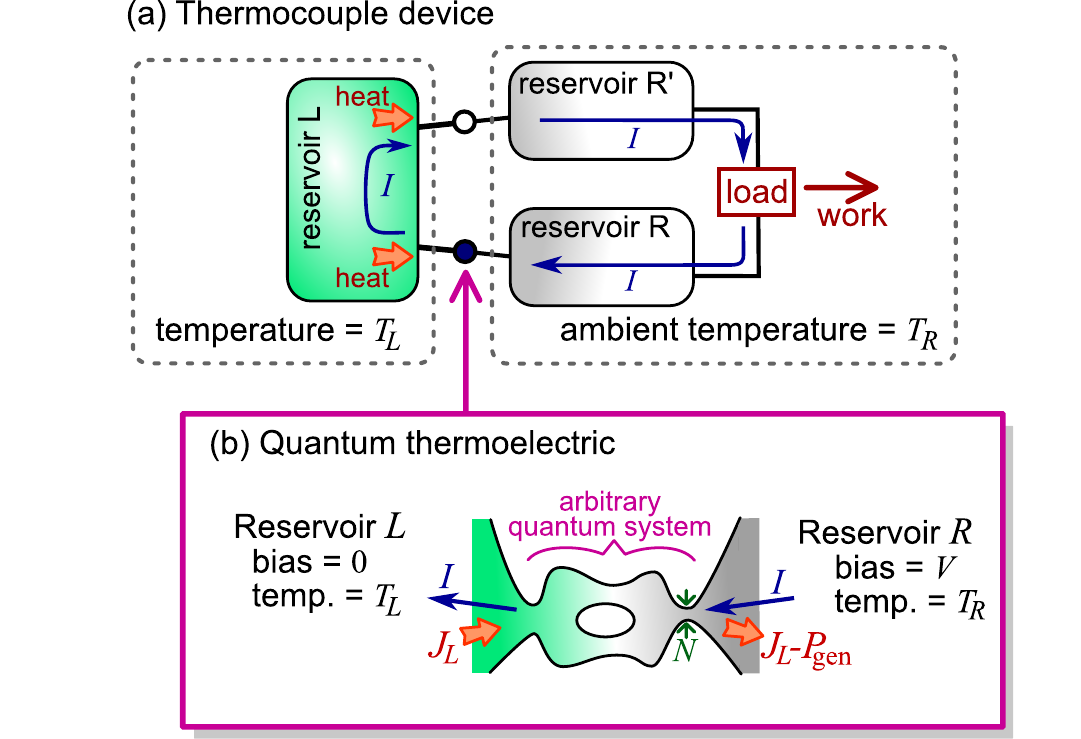}
\caption{\label{Fig:thermocouple} 
(a) The simplest heat-engine is a thermocouple circuit made of two thermoelectrics 
(filled and open circles).
The filled and open circles are quantum systems with opposite thermoelectric responses, 
an example could be that in (b).
For a heat-engine, we assume $T_L > T_R$, so heat flows as shown,
generating a current $I$, which provides power to a load (battery charger, motor, etc.) that converts 
the
electrical power into some other form of work.
The same thermocouple circuit can act as a refrigerator; if one replaces the load with a power supply that generates the current $I$.  This induces the heat flow out of Reservoir $L$, which thereby 
refrigerates Reservoir $L$, so $T_L < T_R$.
Note that in both cases the circuit works because the two
thermoelectrics are electrically in series but thermally in parallel.
In (b), $N$ indicates the number of transverse modes in the narrowest part of the quantum system.
}
\end{figure}

The simplest heat-engine is a thermocouple circuit, as shown in Fig.~\ref{Fig:thermocouple}.
It consists of a pair of thermoelectrics with opposite thermoelectric responses 
(filled and open circles) and a load, connected in a ring.  
Between each such circuit element is a big reservoir of electrons, the reservoir on the left ($L$) 
is hotter than the others, $T_L > T_R$, so heat flows from left to right.
One thermoelectric's response causes an electric current to flow in the opposite direction 
to the heat flow (filled circle), while the other's causes an electric current to flow 
in the same direction as the heat flow (open circle).  
Thus, the two thermoelectrics turn heat energy into electrical work; a current flow $I$ 
through the load.
The load is assumed to be a device that turns the electrical work into some other form of work;
it could be a battery-charger (turning electrical work into chemical work) or a motor (turning electrical work 
into mechanical work).

The same thermocouple circuit can be made into a refrigerator simply by replacing the
load with a power supply.  The power supply does work to establish the current $I$ around the circuit,
 and this current through the thermoelectrics can ``drag'' heat out of reservoir $L$.
In other words, the electrical current and heat flow are the same as for the heat-engine,
but now the former causes the latter rather than vice versa.
Thus, the refrigerator cools reservoir $L$, so $T_L < T_R$. 
 
The laws of classical thermodynamics inform us that entropy production can never be negative, and maximal efficiency occurs
when a system operates reversibly (zero entropy production).  Thus, it places fundamental bounds 
on heat-engine and refrigerator efficiencies, known as Carnot efficiencies.
In both cases, the efficiency is defined as the power output divided by the power input.
For the heat-engine, the power input is the heat current out of the hotter reservoir (reservoir $L$), $J_L$, 
and the power output is the electrical power generated $P_{\rm gen}$.
Thus, the heat-engine (eng) efficiency is 
\begin{eqnarray}
\eta_{\rm eng} = P_{\rm gen}\big/ J_L.
\label{Eq:eff-eng}
\end{eqnarray}  
This efficiency can never exceed Carnot's limit,
\begin{eqnarray}
\eta_{\rm eng}^{\rm Carnot} &=& 1-T_R/ T_L,
\label{Eq:Carnot-eng}
\end{eqnarray}
where we recall that we have $T_L > T_R$.

For the refrigerator the situation is reversed, the load is replaced by a power supply, and the power input is the electrical power that the circuit absorbs from the power supply, $P_{\rm abs}$. The power output is the heat current out of the colder reservoir (reservoir $L$), $J_L$.  This is called the cooling power, because it is the rate at which the circuit removes heat energy from reservoir $L$.
Thus, the refrigerator (fri) efficiency is 
\begin{eqnarray}
\eta_{\rm fri} = J_L \big/ P_{\rm abs}.
\label{Eq:eff-fri}
\end{eqnarray}
This efficiency is often called the coefficient of performance or COP.  
This efficiency can never exceed Carnot's limit,
\begin{eqnarray}
\eta_{\rm fri}^{\rm Carnot} &=& (T_R/T_L -1)^{-1}, \ \  
\label{Eq:Carnot-fri}
\end{eqnarray}
where we recall that $T_L < T_R$ (opposite of heat-engine). 

Strangely, the laws of classical thermodynamics do not appear to place a fundamental bound on the power output associated with reversible (Carnot efficient) operation. Most textbooks say that reversibility requires ``small'' power output, but rarely define what ``small'' means.    
The central objective of Ref.~[\onlinecite{2014w-prl}] was to find the meaning of  ``small'', and find a fundamental upper bound on the efficiency of an irreversible system in which the power output was {\it not} small.

\begin{figure}
\includegraphics[width=\columnwidth]{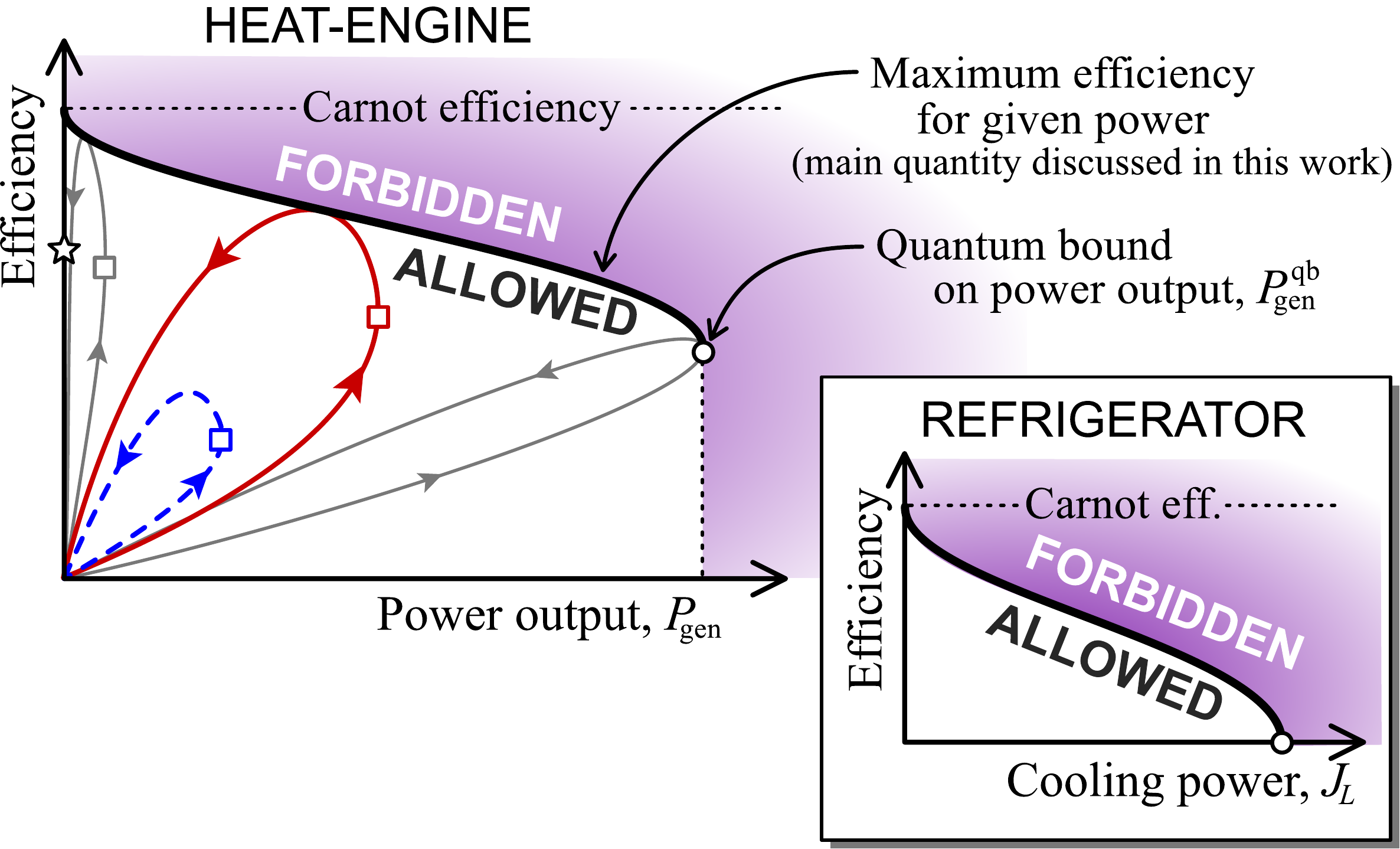}
\caption{\label{Fig:summary} 
The thick black curves are qualitative sketches of the maximum efficiency as a function of 
heat-engine power output (main plot), or refrigerator cooling power (inset),
with the shaded regions being forbidden. 
Precise plot of such curves for different temperature ratios, $T_R/T_L$,
are shown in Fig.~\ref{Fig:allpowers}. 
The colored loops (red, grey and blue) are typical sketches of the efficiency versus power of {\it individual}  heat-engines
as we increase the load resistance (direction of arrows on loop).  
The power output $P_{\rm gen}=IV$ vanishes when the load resistance is 
zero (for which $V=0$) or infinite (for which $I=0$), with a maximum at an intermediate resistance (open square). 
The curves have a characteristic loop form \cite{Casati-review}, 
however the exact shape of the loop depends on many system specific details, such as charging effects.
The dashed blue loop is for a typical non-optimal system (always well below the upper bound), 
while the solid red and grey loops are for systems which achieve the upper bound for a 
particular value of the load.
The star marks the Curzon-Ahlborn efficiency.
}
\end{figure}

Ref.~[\onlinecite{2014w-prl}] did this for the class of quantum thermoelectrics that 
are well modelled by a scattering theory,
which enables one to straightforwardly treat quantum and thermodynamic effects on an equal footing.
It summarized two principal results absent from classical thermodynamics.
Firstly, there is a quantum bound (qb) on the power output, and no quantum system can exceed this bound
(open circles in Fig.~\ref{Fig:summary}).
Secondly, there is a upper bound on the efficiency at any given power output less than this bound
(thick black curves in Fig.~\ref{Fig:summary}).
The efficiency at given power output can only reach Carnot efficiency when the power output
is very small compared to the quantum bound on power output. The upper bound on efficiency then 
decays monotonically as one increases the power output towards the quantum bound.
The objective of this article is to explain in detail the methods used to derive these results,
along with the other results that were summarized in Ref.~[\onlinecite{2014w-prl}].

\subsection{Contents of this article}
This article provides detailed derivations of the results in Ref.~[\onlinecite{2014w-prl}].
The first part of this article is an extended introduction.
Section~\ref{Sect:literature} is a short review of the relevant literature.
Section~\ref{Sect:Unique} discusses how we define temperature, heat and entropy.  
Section~\ref{Sect:entropy-prod} recalls the connection between efficiency and entropy production
in any thermodynamic machine.
Section~\ref{Sect:ScatteringTheory} reviews the nonlinear scattering theory,
which section~\ref{Sect:over-estimates} uses to make very simple over-estimates 
of a quantum system's maximum power output.

The second part of this article considers how to optimize a system
which is free of relaxation and has no phonons or photons.
Section~\ref{Sect:guess-heat} gives a hand-waving explanation of the optimal heat engine,
while Section~\ref{Sect:eng} gives the full derivation. 
Section~\ref{Sect:guess-fri} gives a hand-waving explanation of the optimal refrigerator,
while Section~\ref{Sect:fri} gives the full derivation.
Section~\ref{Sect:chain} proposes a system which could in principle come arbitrarily close to the 
  optimal properties given in sections~\ref{Sect:eng} and \ref{Sect:fri}. 
Section~\ref{Sect:in-parallel} considers many quantum thermoelectrics in parallel.

The third part of this article considers certain effects neglected
in the above idealized system.
Section~\ref{Sect:ph} adds the parasitic effect of phonon or photon carrying heat in parallel to the electrons.
Section~\ref{Sect:Relax} treats relaxation within the quantum system.


\section{Comments on existing literature}
\label{Sect:literature}

There is much interest in using thermoelectric effects to cool fermionic atomic 
gases \cite{Grenier2012,Brantut-Grenier-et-al2013,Grenier2014}, 
which are hard to cool via other methods.  
This physics is extremely similar to that in this work, but there is a crucial difference.
For the electronic systems that we consider, we can assume the temperatures to be much less than the reservoir's Fermi energy, and so take all electrons to have
the same Fermi wavelength.  In contrast, fermionic atomic gases have temperatures of order the Fermi energy, so the high-energy particles in a reservoir have a different wavelength from the low-energy ones. 
Thus, our results do not apply to atomic gases, 
although our methodology does\cite{Grenier2014}.

\subsection{Nonlinear systems and the figure of merit $ZT$}
\label{Sect:nonlinear+ZT}

Engineers commonly state that wide-ranging applications for thermoelectrics would require them to have 
a dimensionless figure of merit,  $ZT$, greater than three. 
This dimensionless figure of merit is a dimensionless combination of the linear-response coefficients \cite{books} $ZT= TGS^2/\Theta$,
for temperature $T$,
 Seebeck coefficient $S$, electrical conductance $G$, and thermal conductance $\Theta$ .
Yet for us, $ZT$ is just a way to characterize the efficiency, via
\begin{eqnarray}
\eta_{\rm eng} = \eta_{\rm eng}^{\rm carnot}  {\sqrt{ZT+1} -1 \over \sqrt{ZT+1} +1}, 
\nonumber
\end{eqnarray}
with a similar relationship for refrigerators.
Thus, someone asking for a device with a $ZT > 3$, 
actually requires one with an efficiency of more than one third of Carnot efficiency.   
This is crucial, because the efficiency is a physical quantity in linear and nonlinear situations, 
while $ZT$ is only meaningful in the linear-response regime \cite{Zebarjadi2007,Grifoni2011,2012w-pointcont,Meair-Jacquod2013,Michelini2014,Azema-Lombardo-Dare2014}. 

Linear-response theory rarely fails for bulk semiconductors, 
even when $T_L$ and $T_R$ are very different. Yet it is completely
{\it inadequate} for the quantum systems that we consider here.
Linear-response theory requires the temperature drop on the scale of 
the electron relaxation length $l_{\rm rel}$
(distance travelled before thermalizing) to be much less than the average temperature.  
For a typical millimetre-thick bulk thermoelectric between a diesel motor's exhaust system 
($T_L\simeq 700$K) and its surroundings ($T_R\simeq 280$K), 
the relaxation length (inelastic scattering length) is of order the mean free path; typically 1-100nm.
The temperature drop on this scale is tens of thousands of times smaller than the temperature drop 
across the whole thermoelectric.  This is absolutely tiny compared with the average temperature,
so linear-response \cite{Mahan-Sofo1996} works well, even though  $(T_L-T_R)/T_L$ is of order one.

In contrast, for quantum systems ($L \ll l_{\rm rel}$), the whole temperature drop occurs on the scale of a few nanometres or less, and so linear-response theory is inapplicable whenever
$(T_L-T_R)/T_L$ is not small.

\subsection{Carnot efficiency}
\label{Sect:Carnot}

A system must be reversible (create no entropy) to have Carnot efficiency;
proposals exist to achieve this in bulk \cite{Mahan-Sofo1996} or quantum 
\cite{Humphrey-Linke2005,Kim-Datta-Lundstrom2009,Jordan-Sothmann-Sanchez-Buttiker2013} thermoelectric. 
It requires that electrons only pass between reservoirs L and R at the energy where 
the occupation probabilities are identical in the two reservoirs \cite{Humphrey-Linke2005}.
Thus, a thermoelectric requires two things to be reversible.
Firstly, it must have a $\delta$-function-like transmission \cite{Mahan-Sofo1996,Humphrey-Linke2005,Kim-Datta-Lundstrom2009,Jordan-Sothmann-Sanchez-Buttiker2013,Sothmann-Sanchez-Jordan-Buttiker2013},
which only lets electrons through at energy $\eps_0$.
Secondly,\cite{Humphrey-Linke2005} the load's resistance must be such  
that $\eminus V = \eps_0 (1-T_R/T_L)$, 
so the reservoirs' occupations are equal at $\eps_0$, see Fig.~\ref{Fig:Fermi}.

By definition this means the current vanishes, and thus so does the power output, $P_{\rm gen}$.
However, one can see how $P_{\rm gen}$ vanishes by 
considering a quantum system which lets electrons through in a tiny energy window $\Delta$ from $\eps_0$ 
to $\eps_0+\Delta$, see Fig~\ref{Fig:tophat-width}.
When we take $\Delta\big/(\kB T_{L,R}) \to 0$, 
one has Carnot efficiency, however we will see
(leading order term in Eq.~(\ref{Eq:Pgen-eng-lowpower})) that
\begin{eqnarray}
P_{\rm gen} \propto {1 \over \hbar} \Delta^2,
\label{Eq:power-for-Delta-to-zero}
\end{eqnarray}
which vanishes as $\Delta\big/(\kB T_{L,R}) \to 0$.

\subsection{Heat-engine efficiency at finite power output and Curzon-Ahlborn efficiency}
\label{Sect:eff-CA}

To increase the power output beyond that of a reversible system,
one has to consider irreversible machines which generate a finite amount of entropy 
per unit of work generated.
Curzon and Ahlborn\cite{Curzon-Ahlborn1975} popularized the idea
of studying the efficiency of a heat-engine running at its maximum power output.
For classical pumps, this efficiency is $\eta_{\rm eng}^{\rm CA} = 1- \sqrt{T_L/T_R}$, which is now called the Curzon-Ahlborn efficiency, 
although already given in Refs.~[\onlinecite{Yvon1956,Chambadal57,Novikov57}].
As refrigerators, these pumps have an efficiency at maximum cooling power of zero, although 
Refs.~[\onlinecite{Velasco1997,Tomas2012,Apertet2013,Correa2014}] discuss ways around this.

The response of a given heat-engine is typically a ``loop'' 
of efficiency versus power (see Fig.~\ref{Fig:summary}) as one varies the 
load on the system\cite{Casati-review}.  For a peaked transmission function with width  $\Delta$
(see e.g.~Fig.~\ref{Fig:tophat-width}), the loop moves to the left as one reduces $\Delta$. 
In the limit $\Delta \to 0$, the whole loop is squashed onto the $P_{\rm gen}=0$ axis.
In linear-response language, this machine has $ZT \to \infty$. 
In this limit,
the efficiency at maximum power can be very close to that of Curzon and Ahlborn \cite{Esposito2009-thermoelec} (the star in Fig.~\ref{Fig:summary}),
just as its maximum efficiency 
can be that of Carnot\cite{Humphrey-Linke2005} (see previous section).
However, its maximum power output is $\propto \eminus V\Delta/\hbar$ for small $\Delta$
(where $V$ is finite, chosen to ensure maximum power), which vanishes for $\Delta \to 0$, although it is much larger than Eq.~(\ref{Eq:power-for-Delta-to-zero}).
Fig.~\ref{Fig:summary} shows that a system with larger $\Delta$ (such as the red curve) operating near its maximum efficiency will have both higher efficiency and higher power output than the one with small $\Delta$ (left most grey curve) operating at maximum power.

This article shows how to derive the thick black curve in 
Fig.~\ref{Fig:summary}, thereby showing that there is a fundamental trade-off between efficiency 
and power output
in optimal thermodynamic machines made from thermoelectrics \cite{footnote:casati-review}.
As such, our work overturns the idea that maximizing efficiency at maximum power is the best route
to machines with both high efficiency and high power.
It also overturns the idea that systems with the 
narrowest transmission distributions (the largest $ZT$ in linear-response)
are automatically the best thermoelectrics. 

At this point we mention that other works\cite{Nakpathomkun-Xu-Linke2010,Leijnse2010,Meden2013,Hershfield2013} have studied efficiencies 
for various systems with finite width transmission functions, for which power outputs can be finite.  
In particular, Ref.~[\onlinecite{Hershfield2013}] considered a boxcar transmission function,
which is the form of transmission function that we have shown can be made optimal \cite{2014w-prl}.

\subsection{Pendry's quantum bound on heat-flow}
\label{Sect:Pendry}

An essential ingredient in this work  
is Pendry's upper bound \cite{Pendry1983} on the heat-flow through a quantum system between two reservoirs of fermions. 
He found this bound using a scattering theory of the type discussed in 
Section~\ref{Sect:ScatteringTheory} below.
It is a concrete example of a general principle due to Bekenstein \cite{Bekenstein},
and the same bound applies
in the presence of thermoelectric effects \cite{2012w-2ndlaw}.
The bound on the heat flow out of reservoir $L$
is achieved when all the electrons and holes arriving at the quantum system from reservoir $L$ escape into reservoir $R$
without impediment, while there is no back-flow of electrons or holes from reservoir $R$ to L.
The easiest way to achieve this is to couple reservoir $L$ through the 
quantum system to a reservoir $R$ at zero temperature, and then ensure the quantum system does not reflect any particles. In this case the heat current equals
\begin{eqnarray}
J^{\rm qb}_L = {\pi^2 \over 6h} N \kB^2 T_L^2,
\label{Eq:Jqb}
\end{eqnarray}
where $N$ is the number of transverse modes in the quantum system.
We refer to this as the quantum bound (qb) on heat flow, 
because it depends on the quantum wave nature of the electrons;
it depends on $N$, which is given by the cross-sectional area of the quantum system 
divided by $\lambda_{\rm F}^2$, where $\lambda_{\rm F}$ is the electron's Fermi wavelength.
As such $J_L^{\rm qb}$ is ill-defined within classical thermodynamics.

\section{Uniquely defining temperature, heat and entropy}
\label{Sect:Unique}

\begin{figure}
\includegraphics[width=\columnwidth]{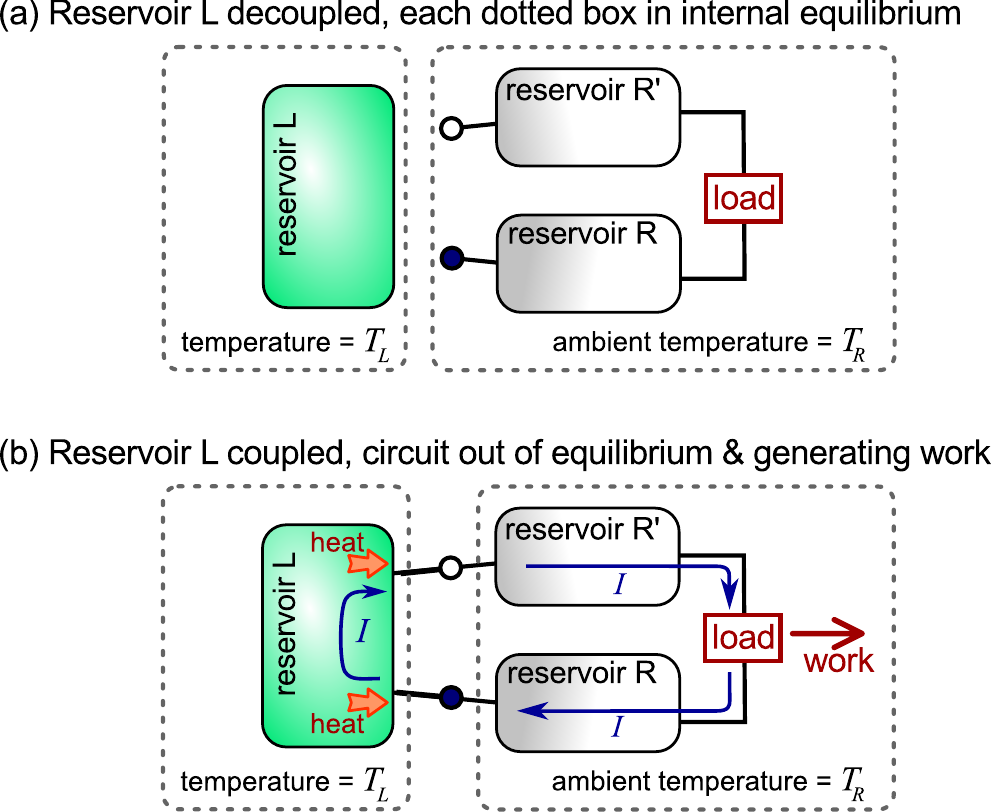}
\caption{\label{Fig:Unique} 
To implement the procedure in Section~\ref{Sect:Unique},
one starts with the circuit unconnected, as in (a), one then connects the circuit, as in (b).
After a long time $t_{\rm expt}$, one disconnects the circuit, returning to (a).
The circles are the quantum thermoelectrics,
as in Fig.~\ref{Fig:thermocouple}.
}
\end{figure}

Works on classical thermodynamics have shown that the definition of heat and entropy flows
can be fraught with difficulties.  For example, the rate of change of entropy cannot always be uniquely defined in classical continuum 
thermodynamics\cite{Kern1975,Day1977,book:irreversible-thermodyn}.
Here the situation is even more difficult, since the
electrons within the quantum systems (circles in 
Fig.~\ref{Fig:thermocouple}) are not at equilibrium, and so their temperature cannot be defined.  
Thus, it is crucial to specify the logic which leads to our definitions of temperature, heat flow and entropy flow.

Our definition of heat flow originated in Refs.~[\onlinecite{Engquist-Anderson1981,Sivan-Imry1986,Butcher1990}],
the rate of change of entropy is then found using the Clausius relation \cite{Footnote:Sivan-Imry} (see below).
To explain these quantities and show they are unambiguous, we 
consider the following three step procedure for a heat engine.
An analogue procedure works for a refrigerator.

\begin{itemize}
\item[]{\bf Step 1.} 
Reservoir $L$ is initially decoupled from the rest of the circuit (see Fig.~\ref{Fig:Unique}a), 
has internal heat energy $Q_L^{(0)}$, and is in internal equilibrium at temperature $T_L^{(0)}$.  
The rest of the circuit is in equilibrium at temperature $T_R^{(0)}$
with internal heat energy $Q_R^{(0)}$. The internal heat energy is 
the total energy of the reservoir's electron gas minus the energy which that gas would have in its ground-state.
As such, the internal energy can be written as a sum over electrons and holes, with an electron at energy $\eps$ above the reservoir's chemical potential (or a hole at energy $\eps$ below that chemical potential) contributing $\eps$ to this internal heat energy.   
The initial entropies are then $S^{(0)}_i = Q^{(0)}_i \big/ T^{(0)}_i$ for $i=L,R$.

\item[]{\bf Step 2.}
We connect reservoir $L$ to the rest of the circuit ( (see Fig.~\ref{Fig:Unique}b)
and leave it connected for a long time $t_{\rm expt}$.
While we assume $t_{\rm expt}$ is long, we also assume that the reservoirs are all large enough that the energy distributions within them change very little during time $t_{\rm expt}$.
Upon connecting the circuit elements, we assume a transient response during a time $t_{\rm trans}$,
after which the circuit achieves a steady-state.  
We ensure that $t_{\rm expt}\gg t_{\rm trans}$,
so the physics is dominated by this steady-state.
Even then the flow will be noisy \cite{Blanter-Buttiker} due to the fact 
electrons are discrete with probabilistic dynamics. 
So we also ensure that $t_{\rm expt}$ is much longer than the noise correlation time, so that the noise in the currents is negligible compared to the average currents.

\item[]{\bf Step 3.}
After the time $t_{\rm expt}$, we disconnect reservoir $L$ from the rest of the circuit. Again, there will be a transient response,  however we assume that a weak relaxation mechanism within the reservoirs will cause the two parts of the circuit to each relax to internal equilibrium (see Fig.~\ref{Fig:Unique}a).
After this one can unambiguously identify the temperature, $T_i$, internal energy $Q_i$ 
and Clausius entropy $S_i=Q_i\big/ T_i$ of the two parts of the circuit (for $i=L,R$).  
Since the reservoirs are large, we assume $T_i = T_i^{(0)}$.
\end{itemize}

Thus, we can unambiguously say that
the heat-current out of reservoir $i$ {\it averaged} over the time $t_{\rm expt}$ is 
\begin{eqnarray}
\langle J_i \rangle = \big (Q^{(0)}_i - Q_i \big) \big/ t_{\rm expt}.
\end{eqnarray}
For the above thermocouple, we treat the currents for each thermoelectric separately,
writing the heat current out of reservoir $L$ as $J_L+J_{L'}$, where
$J_L$ is the heat current from reservoir $L$ into the lower thermoelectric in Fig.~\ref{Fig:thermocouple} 
(the filled circle),
and $J_{L'}$ is the heat current from reservoir $L$ into the upper thermoelectric in Fig.~\ref{Fig:thermocouple} 
(the open circle).
Treating each thermoelectric separately is convenient,
and also allows one to generalize the results to ``thermopiles'',
which contain  hundreds of thermoelectrics arranged so that they are electrically in series, but
thermally in parallel.

The average rate of change of entropy in the circuit is 
$\langle \dot S_{\rm circuit} \rangle = \langle \dot S \rangle +\langle \dot S' \rangle$,
where $\langle \dot S \rangle$ is the average rate of change of entropy associated with the lower 
thermoelectric in Fig.~(\ref{Fig:thermocouple}), while $\langle \dot S' \rangle$ is that for the upper thermoelectric.  Then
\begin{eqnarray}
\langle \dot S\rangle = \langle \dot S_L \rangle + \langle \dot S_R \rangle 
 =  -{\langle J_{\rm L} \rangle \big/ T_L} \,-\, {\langle J_{\rm R} \rangle \big/ T_R}\,,
\label{Eq:average-dotS-def}
\end{eqnarray}
while $\langle \dot S'\rangle$ is the same with $J_L,J_R,T_R$ replaced by $J_{L'},J_{R'},T_{R'}$.
We neglect the entropy of the thermoelectrics and load, by assuming their initial and final state are the same.
This will be the case if they are small compared to the reservoirs, so their initial and final states a simply given by the temperature $T_R$.

The nonlinear scattering theory in Ref.~[\onlinecite{Christen-ButtikerEPL96}] captures
long-time average currents (usually called the DC response in electronics),
such as  electrical current $\langle I_i \rangle$ and heat current $\langle J_i \rangle$,
see references in Section~\ref{Sect:ScatteringTheory}.
It is believed to be exact for non-interacting particles, and
also applies when interactions can be  
treated in a mean-field approximation (see again section \ref{Sect:ScatteringTheory}).
A crucial aspect of the scattering theory is that we do not need to describe the 
non-equilibrium state of the quantum system during step 2.
Instead, we need that quantum system's
transmission function, defined in section \ref{Sect:ScatteringTheory}.

In this article we will {\it only} discuss the long-time average of the rates of flows 
(not the noisy instantaneous flows), and thus 
will not explicitly indicate the average; so $I_i$, $J_i$ and $\dot S_i$ should
be interpreted as  
$\langle I_i \rangle$, $\langle J_i \rangle$ and $\langle\dot S_i \rangle$.

\section{Entropy production}
\label{Sect:entropy-prod}

There are little known universal relations between efficiency, power and and entropy production,
which follow trivially from the laws of thermodynamics \cite{Cleuren2012}.
Consider the lower thermoelectric in Fig.~\ref{Fig:thermocouple}a (filled circle), 
with $J_L$ and $J_R$ being steady-state heat currents into it from reservoir $L$ and R.
Then the first law of thermodynamics is
\begin{eqnarray}
J_R + J_L=P_{\rm gen},
\label{Eq:firstlaw}
\end{eqnarray}
where $P_{\rm gen}$ is the electrical power generated.
The Clausius relation for the
rate of change of total entropy averaged over long times as in Eq.~(\ref{Eq:average-dotS-def}), is
\begin{eqnarray}
\dot S  = -{J_L \over T_L} + {J_L - P_{\rm gen} \over T_R},
\label{Eq:secondlaw}
\end{eqnarray}
where we have used Eq.~(\ref{Eq:firstlaw}) to eliminate $J_R$.

For a heat engine, we take $J_L$ to be positive, which means $T_L > T_R$
and $J_R$ is negative. 
We use Eq.~(\ref{Eq:eff-eng}) to replace $J_L$ with $P_{\rm gen}/\eta_{\rm eng}$ 
in Eq.~(\ref{Eq:secondlaw}).
Then, the rate of entropy production by a heat-engine with efficiency 
$\eta_{\rm eng}(P_{\rm gen})$ at
power output $P_{\rm gen}$ is 
\begin{eqnarray}
\dot S (P_{\rm gen}) 
 &=& 
{P_{\rm gen} \over T_R} \left({\eta_{\rm eng}^{\rm carnot} \over \eta_{\rm eng}(P_{\rm gen})} -1 \right),
\label{Eq:dotS-eng}
\end{eqnarray}
where the Carnot efficiency, $\eta_{\rm eng}^{\rm carnot}$, is given in Eq.~(\ref{Eq:Carnot-eng}).
Hence, knowing the efficiency at power $P_{\rm gen}$, tells us the entropy production at that power.  Maximizing the former minimizes the latter.

For refrigeration, the load in Fig.~\ref{Fig:thermocouple} is replaced by a power supply, 
the thermoelectric thus absorbs a power  $P_{\rm abs}$ to extract heat from the cold reservoir.
We take reservoir $L$  as cold ($T_L < T_R$) , so $J_L$ is positive.
We replace $P_{\rm gen}$ by $-P_{\rm abs}$ in Eqs.~(\ref{Eq:firstlaw},\ref{Eq:secondlaw}). 
We then use  Eq.~(\ref{Eq:eff-fri}) to replace $P_{\rm abs}$ by $J_L/\eta_{\rm fri}$.
Then the rate of entropy production by a refrigerator at cooling power $J_L$ is 
\begin{eqnarray}
\dot S (J_L) 
 &=& 
{J_L \over T_R} \left({1\over \eta_{\rm fri}(J_L)} -{1 \over \eta_{\rm fri}^{\rm carnot}}  \right),
\label{Eq:dotS-fri}
\end{eqnarray}
where the Carnot efficiency, $\eta_{\rm fri}^{\rm carnot}$, is given in Eq.~(\ref{Eq:Carnot-fri}).
Hence knowing a refrigerator's efficiency at cooling power $J_L$ gives us its entropy production,
and we see that maximizing the former minimizes the latter.

Eqs.~(\ref{Eq:dotS-eng},\ref{Eq:dotS-fri}) hold for systems modelled by scattering theory, because 
this theory satisfies the laws of thermodynamics \cite{Bruneau2012}$^,$\cite{2012w-2ndlaw}.
The rate of entropy production is zero when the efficiency is that of Carnot,
but becomes increasingly positive as the efficiency reduces.
In this article, we calculate the maximum efficiency for given power output,
and then use Eqs.~(\ref{Eq:dotS-eng},\ref{Eq:dotS-fri}) to
get the minimum rate of entropy production at that power output.

\section{Nonlinear Scattering Theory}
\label{Sect:ScatteringTheory}

This work uses Christen and B\"uttiker's nonlinear scattering theory \cite{Christen-ButtikerEPL96},
which treats electron-electron interactions as mean-field charging effects.   
Refs.~[\onlinecite{Sanchez-Lopez2013,2012w-pointcont,Meair-Jacquod2013}] added thermoelectric effects by following works on linear-response
\cite{Engquist-Anderson1981,Sivan-Imry1986,Butcher1990}.
Particle and heat flows are given by the transmission function, ${\cal T}_{RL}(\eps)$, 
for electrons to go from left ($L$) to right ($R$) at energy $\eps$, where ${\cal T}_{RL}(\eps)$ is a {\it self-consistently} determined function of $T_L$, $T_R$ and $V$.
In short, this self-consistency condition originates from the fact that electrons injected from the leads change the charge distribution in the quantum system, which in turn changes the behaviour of those 
injected electrons (via electron-electron interactions).
The transmission function can be determined self-consistently with the charge distribution,
if the latter is treated in a time-independent mean-field manner (neglecting single electron effects). 
We note that the same nonlinear scattering theory was also derived for resonant level models 
\cite{Humphrey-Linke2005,Nakpathomkun-Xu-Linke2010} using functional RG to treat single-electron charging effects \cite{Meden2013}. 

The scattering theory for the heat current is based on the observation that
an electron leaving reservoir $i$ at energy $\eps$ is carrying heat $\eps - \mu_i$ out of that reservoir
\cite{Butcher1990}, 
where $\mu_i$ is the reservoir's chemical potential.  Thus, a reservoir is cooled by removing an
electron above the Fermi surface, but heated by removing a electron below the Fermi surface.  
It is convenient to treat empty states below a reference chemical potential 
(which we define as $\eps=0$),
as ``holes''.  Then we do not need to keep track of a full Fermi sea of electrons, but only 
the holes in that Fermi sea.
Then the heat-currents out of reservoirs L and R and into the quantum system 
are
\begin{eqnarray}
J_L  \! &=& \!
{1 \over h} \sum_\mu \int_0^\infty {\rm d}\eps 
\, (\eps -  \mu\eminus V_L) \,  
{\cal T}^{\mu\mu}_{RL}(\eps)  \, \big[f_L^\mu (\eps) - f_R^\mu (\eps)\big],
\nonumber \\
\label{Eq:JL}
\\
J_R  \! &=& \!
{1 \over h} \sum_\mu \int_0^\infty {\rm d}\eps 
\, (\eps -  \mu\eminus V_R) \,  
{\cal T}^{\mu\mu}_{RL}(\eps)   \, \big[f_R^\mu (\eps) - f_L^\mu (\eps)\big],
\nonumber \\
\label{Eq:JR}
\end{eqnarray}
where $\eminus$ is the electron charge ($\eminus<0$),
so $\eminus V_i$ is the chemical potential of reservoir $i$ measured from
the reference chemical potential ($\eps=0$).
The sum is over 
$\mu=1$ for ``electron'' states (full states above the reference chemical potential),
and $\mu=-1$ for ``hole'' states (empty states below that chemical potential).
The Fermi function for particles entering
from reservoir $j$, is 
\begin{eqnarray}
f_j^\mu(\eps) = \left(1+\exp\left[(\eps - \mu \eminus V_j)\big/ (\kB T_j) \right] \right)^{-1}.
\label{Eq:Fermi}
\end{eqnarray}
The transmission function, ${\cal T}^{\nu\mu}_{ij}(\eps)$, is the probability that 
a particle $\mu$ with energy $\eps$ entering the quantum system from reservoir $j$ will
exit into reservoir $i$ as a particle $\nu$ with energy $\eps$.   
We only allow $\nu=\mu$ here, since we do not consider electron to hole scattering within the quantum system (only common when superconductors are present).
Interactions mean that ${\cal T}^{\mu\mu}_{RL}(\eps)$, 
 is a {\it self-consistently} determined function of  $T_L$, $T_R$ $V_L$ and $V_R$.
 
The system generates power $P_{\rm gen} = (V_R-V_L) I_L$, 
so
\begin{eqnarray}
P_{\rm gen} \! &=& \!  
{1\over h} \sum_\mu \int_0^\infty {\rm d}\eps 
\ \mu\eminus (V_R-V_L)\,
{\cal T}^{\mu\mu}_{RL}(\eps) \, \big[f_L^\mu (\eps) - f_R^\mu (\eps)\big], 
\nonumber \\  
\label{Eq:Pgen}
\end{eqnarray}
It is easy to verify that Eqs.~(\ref{Eq:JL}-\ref{Eq:Pgen}) 
satisfy the first law of thermodynamics, Eq.~(\ref{Eq:firstlaw}).
This theory assumes the quantum system to be relaxation-free, 
although decoherence is allowed as it does not change the structure of Eqs.~(\ref{Eq:JL}-\ref{Eq:Pgen}).
Relaxation is discussed in Section~\ref{Sect:Relax}.

We define the voltage drop as $V=V_R-V_L$. Without loss of generality we take the reference chemical potential to be that of reservoir $L$, so
\begin{eqnarray}
V_L=0, \qquad V_R=V,
\label{Eq:def-V}
\end{eqnarray}
then $J_L$ and $P_{\rm gen}$  coincide with Eqs.~(8,9) in Ref.~[\onlinecite{2014w-prl}].

Numerous works have found the properties of thermoelectric systems 
from their transmission functions, ${\cal T}_{RL}(\eps)$.  Linear-response examples include
Refs.~[\onlinecite{Engquist-Anderson1981,Sivan-Imry1986,Butcher1990,Molenkamp1992,Paulsson-Datta2003,Vavilov-Stone2005,
Nozaki2010,jw-epl,Casati2011,Sanchez-Serra2011,Saha2011,Karlstrom2011,jwmb,Hwang-Lopez-Lee-Sanchez2014,Sothmann-Nernst-engine,Linke2013-onsager}],
while nonlinear responses were considered in
Refs.~[\onlinecite{
Galperin2007-2008,
Murphy2008, 
Nakpathomkun-Xu-Linke2010,
Sanchez-Lopez2013,2012w-pointcont,Meair-Jacquod2013,Meden2013,Linke2013b,Battista2014,Sierra-Sanchez2014}],
see Refs.~[\onlinecite{Casati-review,Sothmann-Sanchez-Jordan-review,Haupt-review}]
for recent reviews.
However, here we do not  ask what is the efficiency of a given system,
we ask what is the system that would achieve the highest efficiency, and what is this efficiency?
This is similar in spirit to Ref.~[\onlinecite{Mahan-Sofo1996}], except that we maximize the efficiency
for given power output.

We need to answer this question in the context of the mean-field treatment of electron-electron interactions\cite{Christen-ButtikerEPL96}, in which the transmission function for any given system is the solution of the above mentioned self-consistency procedure.
Despite this complexity, any transmission function (including all mean-field 
interactions) must obey 
\begin{eqnarray}
0\leq{\cal T}^{\mu\mu}_{RL}(\eps)\leq N \ \ \hbox{ for all }\eps,
\label{Eq:basic-limits-on-transmisson}
\end{eqnarray}
where $N$ is the number of transverse modes at the narrowest point in the nanostructure, 
see Fig.~\ref{Fig:thermocouple}.
Let us assume that this is the {\it only} constraint on the transmission function.
Let us assume that for any given $T_L$, $T_R$ and $V$,
a clever physicist could engineer any desired transmission function, so long as it
obeys Eq.~(\ref{Eq:basic-limits-on-transmisson}).  Presumably they could do this
either by solving the self-consistency equations for  ${\cal T}^{\mu\mu}_{RL}(\eps)$, 
or by experimental trial and error.  
Thus, in this work, we find the ${\cal T}^{\mu\mu}_{RL}(\eps)$ 
which maximizes the efficiency given solely the constraint 
in Eq.~(\ref{Eq:basic-limits-on-transmisson}),
and get this maximum efficiency.  
We then rely on future physicists to find a way to construct a system with this 
${\cal T}^{\mu\mu}_{RL}(\eps)$ (although some hints are given in Section~\ref{Sect:chain}).

\section{From thermoelectric optimization
to thermocouple optimization}
\label{Sect:transforming-from-full-to-open}

The rest of this article considers optimizing a single thermoelectric. 
However, an optimal thermocouple heat engine (or refrigerator) consists of two systems with opposite thermoelectric responses (full and open circles in  Fig.~\ref{Fig:thermocouple}).
So here we explain how to get the optimal thermocouple from the optimal thermoelectric.

Suppose the optimal system between $L$ and $R$ (the full circle) 
has a given transmission function 
${\cal T}_{RL}^{\mu,\mu} (\eps)$, which we will find in Section~\ref{Sect:eng}.
This system generates an electron flow parallel to heat flow
(so electric current is anti-parallel to heat flow, implying a negative Peltier coefficient).
The system between $L$ and $R'$ (the open circle) must have the opposite response.
For this we interchange the role played by electrons and holes compared with 
${\cal T}_{RL}^{\mu,\mu} (\eps)$, so the optimal system between $L$ and $R'$ has
\begin{eqnarray}
{\cal T}_{R'L}^{\mu,\mu} (\eps) &=& {\cal T}_{RL}^{-\mu,-\mu} (\eps).
\end{eqnarray}
If the optimal bias for the system between $L$ and $R$ is $V$ (which we will also find in Section~\ref{Sect:eng}), then the optimal bias for the system between $L$ and $R'$ is $-V$.
Then the heat flow from reservoir $L$ into $R'$ equals that from $L$ into $R$, 
while the electrical current from $L$ into $R'$ is opposite to that from $L$ into $R$, 
and so $P_{\rm gen}$ is the same for each thermoelectric.  
The load across the thermocouple (the two thermoelectrics) must be chosen such that
the bias across the thermocouple  is $2V$.  The condition that the charge current out of $L$ equals that into $L$ will then ensure that both thermoelectrics are at their optimal bias.

In the rest of this article we discuss power output and heat input {\it per thermoelectric}.
For a thermocouple, one simply needs to multiply these by
two, so the efficiency is unchanged but the power output is doubled.

\section{Simple estimate of bounds on power output}
\label{Sect:over-estimates}

One of the principal results of Ref.~[\onlinecite{2014w-prl}]
is the quantum bounds on the power output of heat-engines and refrigerators.
The exact derivation of these bounds is given in Sections~\ref{Sect:qb-eng}
and \ref{Sect:qb-fri}.  Here, we give simple arguments for their basic form
based on Pendry's limit of heat flow discussed in Section~\ref{Sect:Pendry} above.

For a refrigerator, it is natural to argue that the upper bound on cooling power will be closely related to Pendry's bound, Eq.~(\ref{Eq:Jqb}).
We will show in Section~\ref{Sect:qb-fri} that this is the case. 
A two-lead thermoelectric can extract as much as half of $J^{\rm qb}_L$.
In other words, the cooling power of any refrigerator must obey
\begin{eqnarray}
J_L &\leq& {1 \over 2} J^{\rm qb}_L \ =\ {\pi^2 \over 12h} N \kB^2 T_L^2.
\end{eqnarray}

Now let us turn to a heat-engine operating between a hot reservoir $L$ and cold reservoir $R$.
Following Pendry's logic, we can expect that the heat current into the quantum system from reservoir $L$ cannot be 
more than $J_L^{\hbox{\scriptsize over-estimate}}  ={\pi^2 \over 6h} N \kB^2 (T_L^2-T_R^2)$.  
Similarly, no heat engine 
can exceed Carnot's efficiency, Eq.~(\ref{Eq:Carnot-eng}).  
Thus, we can safely assume any system's power output is less than 
\begin{eqnarray}
P_{\rm gen}^{\hbox{\scriptsize over-estimate}} 
&=& \eta_{\rm eng}^{\rm carnot} J_L^{\hbox{\scriptsize over-estimate}}  
\nonumber \\
&=&  {\pi^2 N \kB^2 (T_L+T_R) (T_L-T_R)^2 \over 6h \ T_L} .
\end{eqnarray}
We know this is a significant over-estimate, because maximal heat flow cannot coincide with Carnot efficiency.  Maximum heat flow requires ${\cal T}^{\mu\mu}_{RL}(\eps)$ is maximal for all $\eps$ and $\mu$, while
Carnot efficiency requires a ${\cal T}^{\mu\mu}_{RL}(\eps)$ with a 
$\delta$-function-like dependence on $\eps$ (see Section~\ref{Sect:Carnot}). 
None the less, the full calculation in Section~\ref{Sect:qb-eng} shows that the true quantum bound on
power output is such that \cite{footnote:qb2}
\begin{eqnarray}
P_{\rm gen} &\leq& P_{\rm gen}^{\rm qb2} \,\equiv\, 
 A_0\, {\pi^2 \over h} N \kB^2 \big(T_L-T_R\big)^2, \quad \quad
\end{eqnarray}
where $A_0 \simeq 0.0321$. 
Thus, the simple over-estimate of the bound, 
$P_{\rm gen}^{\hbox{\scriptsize over-estimate}}$,   
differs from the true bound $P_{\rm gen}^{\rm qb2}$ by a factor of  
$(1+T_R/T_L)/(6A_0)$. In other words it over estimates the quantum bound by a factor between 5.19 and 10.38 (that is 5.19 when
$T_R=0$ and 10.38 when $T_R=T_L$).
This is not bad for such a simple estimate.


\begin{figure}
\includegraphics[width=0.9\columnwidth]{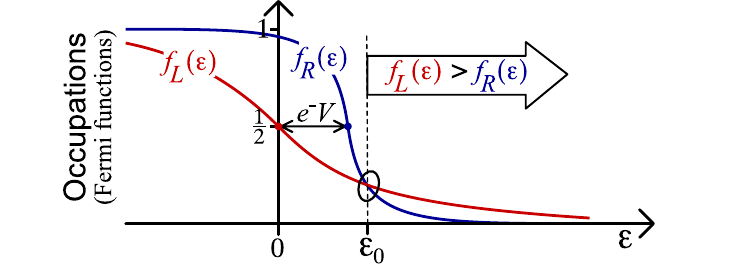}
\caption{\label{Fig:Fermi} 
Sketch of Fermi functions $f_L^\mu(\eps)$ and  $f_L^\mu(\eps)$ 
in Eq.~(\ref{Eq:Fermi}), when $\mu \eminus V$ is positive,
and $T_L > T_R$.  Eq.~(\ref{Eq:Eps0-guess}) gives the point where the two curves cross, $\eps_0$.}
\end{figure}

\section{Guessing the optimal transmission for a heat-engine}
\label{Sect:guess-heat}
Here we use simple arguments to guess the
transmission function which will maximize a heat-engine's efficiency for a given power output.
We consider the flow of electrons 
from reservoir $L$ to reservoir $R$ (the filled circle Fig.~\ref{Fig:thermocouple}a, 
remembering $\eminus<0$, so electron flow is in the opposite direction to $I$).
To produce power, the electrical current must flow against a bias, so we require $\eminus V$ to be positive,
with $V$ as in Eq.~(\ref{Eq:def-V}).
Inspection of the integrand of Eq.~(\ref{Eq:Pgen})
shows that it only gives positive contributions to the power 
output, $P_{\rm gen}$, when $\mu \big(f^\mu_L(\eps) - f^\mu_R(\eps)\big) >0$.
From Eq.~(\ref{Eq:Fermi}),
one can show that $f^\mu_L(\eps)$ and $f^\mu_R(\eps)$ cross at 
\begin{eqnarray}
\eps_0 = \mu \eminus V \big/ (1-T_R/T_L),
\label{Eq:Eps0-guess}
\end{eqnarray}
see Fig.~\ref{Fig:Fermi}.
Since $\eminus V$ is positive, we maximize the power output by blocking the transmission of 
those electrons ($\mu=1$) which have  $\eps< \eps_0$, and blocking the transmission all holes ($\mu=-1$). 
For $\mu=1$, all energies above $\eps_0$ add to the power output. 
Hence, maximizing transmission for all $\eps > \eps_0$
will maximize the power output, giving $P_{\rm gen}=P_{\rm gen}^{\rm qb}$.
However, a detailed calculation, such as that in Section~\ref{Sect:eng},
is required to find the $V$ which will maximize $P_{\rm gen}$; remembering that $P_{\rm gen}$ 
depends directly on $V$ as well as indirectly (via the above choice of $\eps_0$).

Now we consider maximizing the efficiency at a given power output $P_{\rm gen}$, where 
$P_{\rm gen} < P_{\rm gen}^{\rm qb}$.
Comparing the integrands 
in Eqs.~(\ref{Eq:JL},\ref{Eq:Pgen}), we see that $J_L$ contains an extra factor of energy $\eps$
compared to $P_{\rm gen}$.  As a result, the transmission of electrons ($\mu=1$) with large $\eps$ enhances the heat current much more than it enhances the power output. This means that the higher an electron's $\eps$ is,  
the less efficiently it contributes to power production. 
Thus, one would guess that it is optimal to have an upper cut-off on 
 transmission, $\eps_1$, which would be just high enough to ensure the 
 desired power output $P_{\rm gen}$, but no higher.  Then the transmission function will look like a 
``band-pass filter''  (the ``boxcar'' form in Fig~\ref{Fig:tophat-width}),
with $\eps_0$ and $\eps_1$ further apart for higher power outputs.  This guess is correct,
however the choice of $V$ affects both $\eps_0$ and $\eps_1$, so the
calculation in Section~\ref{Sect:eng} is necessary to find the $V$, $\eps_0$ and $\eps_1$ which 
maximize the efficiency for given $P_{\rm gen}$.


\section{Maximizing heat-engine efficiency
for given power output}
\label{Sect:eng}

\begin{figure}[t]
\includegraphics[width=\columnwidth]{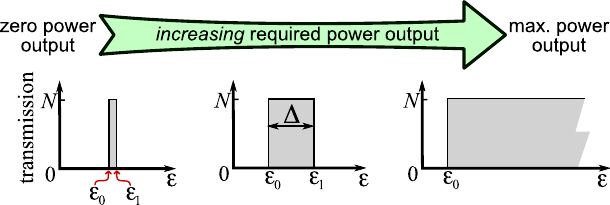}
\caption{\label{Fig:tophat-width} 
How the optimal ``boxcar'' transmission changes with increasing required power output.
At maximum power output, a heat engine has $\eps_1 = \infty$ while $\eps_0$ remains finite.
At maximum cooling power, a refrigerator has $\eps_1 = \infty$ and $\eps_0=0$.
The qualitative features follow this sketch for all $T_R/T_L$, 
however the details depend on
$T_R/T_L$, see Fig.~\ref{Fig:Delta+V}. 
}
\end{figure}

Now we present the central calculations of this article,
finding the maximum efficiency of a quantum thermoelectric
with {\it given} power output.
In this section we consider heat-engines, while Section~\ref{Sect:fri} addresses refrigerators.

\begin{figure}
\includegraphics[width=0.85\columnwidth]{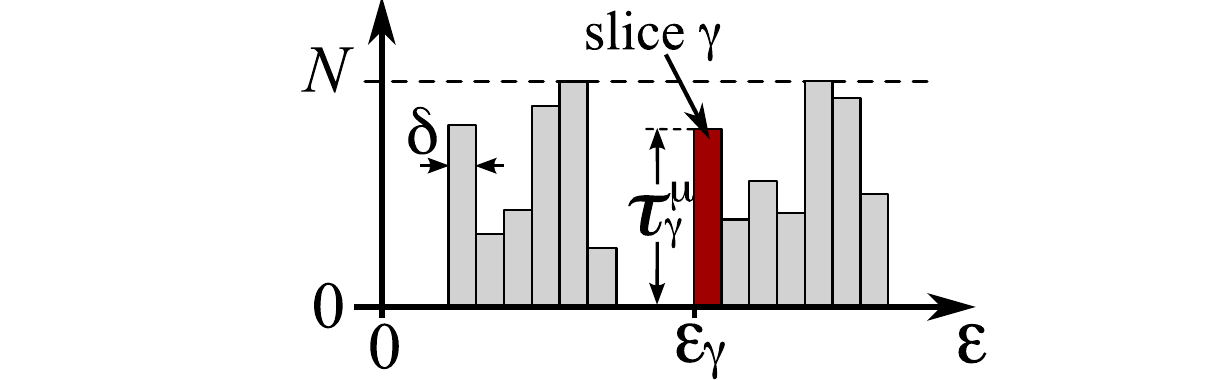}
\caption{\label{Fig:T-functions}
A completely arbitrary transmission function $ {\cal T}_{RL}^{\mu\mu} (\eps)$ 
(see Section \ref{Sect:eng}). We take it to have  
infinitely many slices of width $\delta \to 0$, so slice $\gamma$ has energy 
$\eps_\gamma \equiv \gamma \delta$ and height $\tau^{\mu}_\gamma$. We find the optimal height for each slice. 
}
\end{figure}

For a heat engine, our objective is to find
the transmission function, ${\cal T}^{\mu\mu}_{RL}(\eps)$, and bias, $V$,
that maximize the efficiency $\eta_{\rm eng}(P_{\rm gen})$ for given power output $P_{\rm gen}$.   
To do this we treat ${\cal T}^{\mu\mu}_{RL}(\eps)$
as a set of many slices each of width $\delta \to 0$, see the sketch in Fig.~\ref{Fig:T-functions}a.
We define $\tau^{\mu}_\gamma$ as the height of the $\gamma$th slice, which is at energy 
$\eps_\gamma \equiv \gamma\delta$.
Our objective is to find the optimal value of $\tau^{\mu}_\gamma$ for each $\mu,\gamma$, 
and optimal values of the bias, $V$; all
under the constraint of fixed $P_{\rm gen}$.
Often such optimization problems are formidable,
however this one is fairly straightforward.

The efficiency is maximum for a fixed power, $P_{\rm gen}$,  if 
$J_L$ is minimum for that $P_{\rm gen}$.
If we make an infinitesimal change of $\tau^{\mu}_\gamma$ and $V$,
we note that 
\begin{eqnarray}
\delta P_{\rm gen} &=& \left. {\partial P_{\rm gen} \over \partial \tau^{\mu}_\gamma} \right|_V   \delta\tau^{\mu}_\gamma \ +\ P'_{\rm gen} \,\delta V,
\label{Eq:deltaPgen}
\\
\delta J_L &=& \left. {\partial J_L \over \partial \tau^{\mu}_\gamma} \right|_V   \delta\tau^{\mu}_\gamma \ +\ J'_L \,\delta V,
\label{Eq:deltaJL}
\end{eqnarray} 
where $|_x$ indicates that the derivative is taken at constant $x$,
and the primed indicates $\partial/\partial V$ for fixed transmission functions.
If we want to fix $P_{\rm gen}$ as we change $\tau^{\mu}_\gamma$, we must change the bias $V$ to compensate.  
For this, we set $\delta P_{\rm gen}=0$ in Eq.~(\ref{Eq:deltaJL}) and substitute the result for $\delta V$ into 
Eq.~(\ref{Eq:deltaPgen}).   
Then $J_L$ decreases (increasing efficiency)
for an infinitesimal increase of $\tau^{\mu}_\gamma$  at fixed $P_{\rm gen}$, if
\begin{eqnarray}
\left.{\partial J_L \over \partial \tau^{\mu}_\gamma} \right|_{P_{\rm gen}} 
&=&
\left.{\partial J_L \over \partial \tau^{\mu}_\gamma} \right|_V  
-  {J'_L \over P'_{\rm gen}} 
\left.{\partial P_{\rm gen} \over \partial \tau^{\mu}_\gamma }\right|_V   \ <\ 0.
\qquad \label{Eq:eng-condition}
\end{eqnarray}
Comparing 
Eq.~(\ref{Eq:JL}) and Eq.~(\ref{Eq:Pgen}), one sees that
\begin{eqnarray}
\left.{\partial J_L\over \partial \tau^{\mu}_\gamma }\right|_V &=&
{\eps_\gamma\over \mu \eminus V}\,
\left.{\partial P_{\rm gen}  \over \partial \tau^{\mu}_\gamma }\right|_V  .
\label{Eq:change-JL-to-change-Pgen}
\end{eqnarray}
where $V$ is given in Eq.~(\ref{Eq:def-V}).
Thus, the efficiency $\eta_{\rm eng}(P_{\rm gen})$ grows with a small increase of $\tau^{\mu}_\gamma$ if 
\begin{eqnarray}  
\left(\eps_\gamma - \mu \eminus V  {J'_L \over P'_{\rm gen}} \right) \times
\left.{\partial P_{\rm gen} \over \partial \tau^{\mu}_\gamma }\right|_V   \ <\ 0,
\label{Eq:eng-condition2}
\end{eqnarray}
where $P_{\rm gen}$, $P'_{\rm gen}$, $J_L$, $J'_L$ and $\eminus V$ are positive.

For what follows, let us define 
two energies
\begin{eqnarray}
\eps_0 &=& \eminus V  \big/ (1-T_R/T_L), 
\label{Eq:eng-bounds-eps0}
\\
\eps_1 &=&  \eminus V \,  J'_L / P'_{\rm gen}.
\label{Eq:eng-bounds-eps1}
\end{eqnarray}
One can see that
$ \left.\left({\partial P_{\rm gen}/\partial\tau^{\mu}_\gamma }\right)\right|_V >0$
when both $\mu=1$ and $\eps > \eps_0$, and is negative otherwise. 
Thus, for $\mu=1$,  Eq.~(\ref{Eq:eng-condition2}) is satisfied when $\eps_\gamma$ is between
$\eps_0$ and $\eps_1$.  For $\mu=-1$, Eq.~(\ref{Eq:eng-condition2}) is never satisfied.

A heat-engine is only useful if $P_{\rm gen}>0$, and this is only true for  
$\eps_0 <\eps_1$.
Hence, if $\mu=1$ and $\eps_0 <\eps<\eps_1$, then 
$\eta_{\rm eng}(P_{\rm gen})$ is maximum for  $\tau^{\mu}_\gamma$ at its maximum value,
$\tau^{\mu}_\gamma=N$.  
For all other $\mu$ and $\eps_\gamma$, $\eta_{\rm eng}(P_{\rm gen})$ is maximum 
for  $\tau^{\mu}_\gamma$ at its minimum value,
$\tau^{\mu}_\gamma=0$. 
Since the left-hand-side of 
Eq.~(\ref{Eq:eng-condition2}) is not zero for any $\eps_\gamma\neq \eps_0,\eps_1$, 
there are no stationary points, which is why 
$\tau^{\mu}_\gamma$ never takes a value between its maximum and minimum values.
Thus, the optimal ${\cal T}^{\mu\mu}_{RL}(\eps)$ is a ``boxcar'' or ``top-hat'' function,
\begin{eqnarray}
{\cal T}^{\mu\mu}_{RL}(\eps)
\! &=& \! \left\{ \! \begin{array}{cl} 
N & \hbox{ for }  \mu=1 \ \hbox{ \& } \  \  \eps_0 \! <\! \eps \! 
<\! \eps_1  \phantom{\big|}   
\\
0 & \hbox{ otherwise }  \phantom{\big|} \end{array} \right. \quad 
\label{Eq:top-hat}
\end{eqnarray} 
see Fig.~\ref{Fig:T-functions}b. 
It hence acts as a band-pass filter, only allowing flow between L and R for electrons ($\mu=1$) in the energy window between 
$\eps_0$ to $\eps_1$.

Substituting a boxcar transmission function with arbitrary $\eps_0$ and $\eps_1$ into 
Eqs.~(\ref{Eq:JL},\ref{Eq:Pgen}) gives
\begin{eqnarray}
J_L &=& N \,
\big[F_L(\eps_0)-F_R(\eps_0)-F_L(\eps_1)+F_R(\eps_1) \big],
\label{Eq:JL-eng}
\\ 
P_{\rm gen} \!\! &=& \!N\eminus V \,  \big[G_L(\eps_0)-G_R(\eps_0)
-G_L(\eps_1)+G_R(\eps_1) \big],  \qquad
\label{Eq:Pgen-eng}
\end{eqnarray}
where we define
\begin{eqnarray}
F_j(\eps) = {1 \over h} \int_\eps^\infty 
{ x \ \rmd x \over 
1+ \exp\big[(x-\eminus V_j)\big/(\kB T_j)\big] },
\label{Eq:Fintegral}
\\
G_j(\eps) = {1 \over h}\int_\eps^\infty 
{\rmd x  \over 
1+ \exp\big[(x-\eminus V_j)\big/(\kB T_j)\big] },
\label{Eq:Gintegral}
\end{eqnarray}
which are both positive for any $\eps>0$.
Remembering that we took $V_L=0$ and $V_R=V$,
these integrals are
\begin{eqnarray}
F_L(\eps) &=& \eps G_L(\eps) -{(\kB T_L)^2 \over h} 
{\rm Li}_2\big[-\e^{-\eps/(\kB T_L)}\big], \qquad \\
F_R(\eps) &=& \eps G_R(\eps) -{(\kB T_R)^2 \over h} 
{\rm Li}_2\big[-\e^{-( \eps-\eminus V)/(\kB T_R)}\big], \qquad \\
G_L(\eps) &=& {\kB T_L\over h}\ln\big[1+\e^{-\eps/(\kB T_L)}\big],
\\
G_R(\eps) &=& {\kB T_R\over h}\ln\big[1+\e^{-(\eps-\eminus V)/(\kB T_R)}\big],
\end{eqnarray}
for dilogarithm function, 
${\rm Li}_2(z)= \int_0^\infty t \, dt \big/(\e^t/z -1)$.

We are only interested in cases where $\eps_0$ fulfills the condition in Eq.~(\ref{Eq:eng-bounds-eps0}),
in this case $(\eps_0-\eminus V)/(\kB T_R) = \eps_0/(\kB T_L)$, which means $G_R(\eps_0)$ and $F_R(\eps_0)$ are related to $G_L(\eps_0)$ and $F_L(\eps_0)$ by
\begin{eqnarray}
G_R(\eps_0)&=&{T_R \over T_L}\, G_L(\eps_0),
\label{Eq:G_R}
\\
F_R(\eps_0)-\eps_0 G_R(\eps_0)&=&{T_R^2\over T_L^2}\, \left( F_L(\eps_0) -\eps_0 G_L(\eps_0) \right). \quad
\end{eqnarray}

\begin{figure}
\includegraphics[width=\columnwidth]{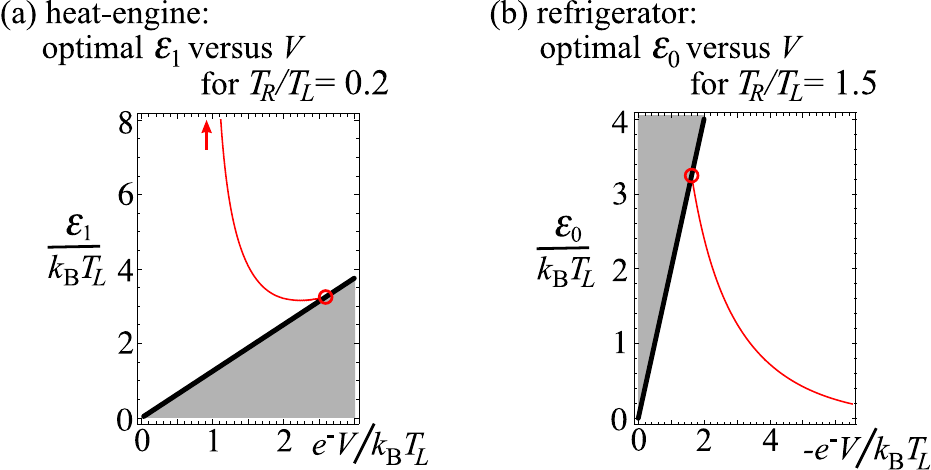}
\caption{\label{Fig:bounds} 
Solutions of the transcendental equations giving optimal 
$\eps_1$ (heat-engine) or  $\eps_0$ (refrigerator).
In (a), the red curve is the optimal $\eps_1(V)$ for $\eps_1> \eps_0$, 
and the thick black line is $\eps_0$ in Eq.~(\ref{Eq:eng-bounds-eps0}). 
The red circle and red arrow indicate the low and high power limits discussed in the text.
In (b), the red curve is the optimal $\eps_0(V)$ for $\eps_0<\eps_1$, and the thick black line
is $\eps_1$ in Eq.~(\ref{Eq:fri-bounds-eps1}).  }
\end{figure}

Eq.~(\ref{Eq:eng-bounds-eps1}) 
tells us that $\eps_1$ depends on $J_L$ and $P_{\rm gen}$, 
but that these depend in-turn on $\eps_1$.
Hence to find $\eps_1$, we substitutes Eqs.~(\ref{Eq:JL-eng},\ref{Eq:Pgen-eng})  into
Eq.~(\ref{Eq:eng-bounds-eps1}) to get a transcendental equation for
$\eps_1$ as a function of $V$ for given $T_R/T_L$.
This equation is too hard to solve analytically
(except in the high and low power limits, discussed in Sections \ref{Sect:qb-eng} and  \ref{Sect:eff-at-given-power} respectively).
The red curve in Fig.~\ref{Fig:bounds}a is 
a numerical solution for $T_R/T_L=0.2$. 

Having found $\eps_1$ as a function of $V$ for given $T_R/T_L$,
we can use  Eqs.~(\ref{Eq:JL-eng},\ref{Eq:Pgen-eng}) to get
$J_L(V)$ and $P_{\rm gen}(V)$.  We can then invert the second relation to get
$V(P_{\rm gen})$.  At this point we can find $J_L(P_{\rm gen})$,
and then use Eq.~(\ref{Eq:eff-eng}) to get the quantity that we desire --- the
maximum efficiency at given power output, $\eta_{\rm eng}(P_{\rm gen})$.

In Section~\ref{Sect:qb-eng}, we do this procedure analytically
for  high power ($P_{\rm gen} =P_{\rm gen}^{\rm qb2}$),
and in Section~\ref{Sect:eff-at-given-power}, we do this procedure analytically
for low power ($P_{\rm gen} \ll P_{\rm gen}^{\rm qb2}$).
For other cases, we only have a numerical
solution for the  transcendental equation for
$\eps_1$ as a function of $V,T_R/T_L$, so we must do everything numerically.

\begin{figure}
\includegraphics[width=\columnwidth]{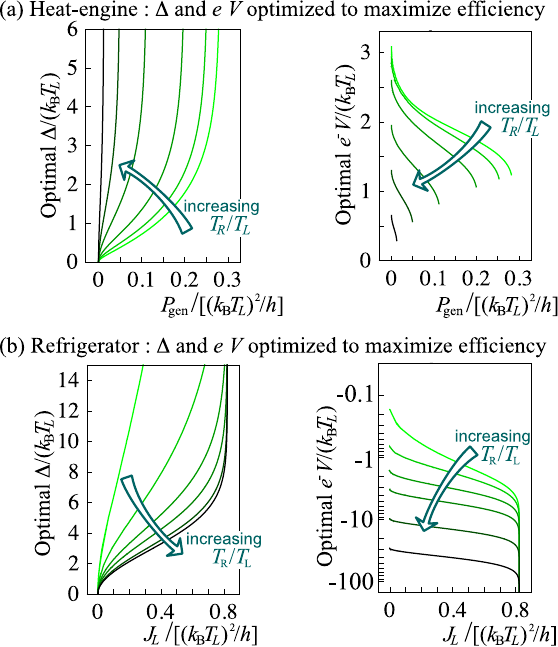}
\caption{\label{Fig:Delta+V}
(a) Plots of optimal $\Delta$ (left) and $\eminus V$ (right)
for a heat-engine with given power output, $P_{\rm gen}$,
for $T_R/T_L=$ 0.05, 0.1, 0.2, 0.4, 0.6 and 0.8.
We get $\eps_0$ from $\eminus V$ by using Eq.~(\ref{Eq:eng-bounds-eps0}).
(b) Plots of optimal  $\Delta$ (left) and $\eminus V$ (right) for a refrigerator 
with a given cooling power output, $J_L$, for $T_R/T_L=$ 1.05, 1.2, 1.5, 2, 4 and 10.
We get $\eps_1$ from $\eminus V$ by using 
Eq.~(\ref{Eq:fri-bounds-eps1}).
}
\end{figure}

\begin{figure}
\includegraphics[width=\columnwidth]{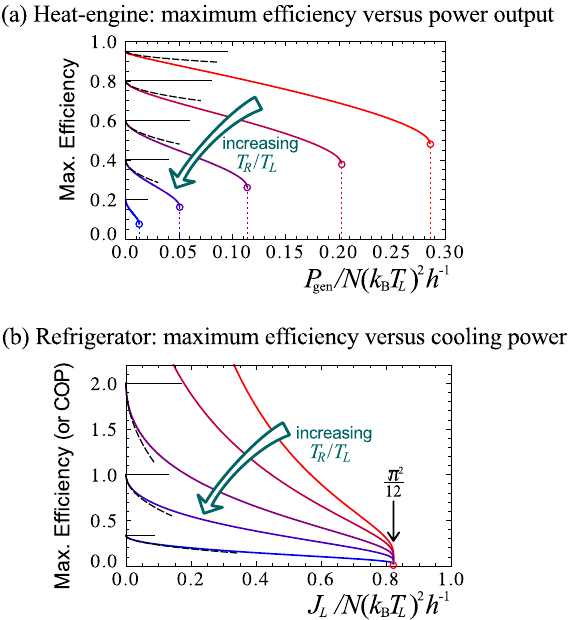}
\caption{\label{Fig:allpowers}
Efficiencies of (a) heat-engines and (b) refrigerators. 
In (a) the curves are the maximum allowed heat-engine efficiency as a function of
power outputs for $T_R/T_L= 0.05,0.2,0.4,0.6,0.8$ (from top to bottom).  
In (b) the curves are the maximum allowed refrigerator efficiency as a function of
cooling power for $T_R/T_L= 1.05,1.2,1.5,2,4$ (from top to bottom).  
In both (a) and (b) the horizontal black lines indicate Carnot efficiency for each $T_R/T_L$,
while the dashed black curves are the analytic theory for small cooling power,
given in Eq.~(\ref{Eq:eta-eng-small-Pgen}) or Eq.~(\ref{Eq:eta-fri-smallJ}).
The circles mark the analytic result for maximum power output. 
}
\end{figure}

Fig.~\ref{Fig:Delta+V}a gives the values of $\Delta=(\eps_1-\eps_0)$ 
and $\eminus V$ which result from solving the transcendental equation numerically
for a variety of different $T_R/T_L$. Eq.~(\ref{Eq:eng-bounds-eps0}) then relates
$\eps_0$ to $\eminus V$.
The qualitative behaviour of the resulting boxcar transmission function is shown in Fig.~\ref{Fig:tophat-width}.
This numerical evaluation enables us to find the efficiency as a function of 
$P_{\rm gen}$ and $T_R/T_L$, which we  
plot in Fig.~\ref{Fig:allpowers}a.

\subsection{Quantum bound on heat engine power output}
\label{Sect:qb-eng}

Here we want to find the highest possible power output of the heat-engine.
In the previous section, 
we had the power as a function of two independent parameters,
$V$ and $\eps_1$, with $\eps_0$ given by Eq.~(\ref{Eq:eng-bounds-eps0}). 
However, we know that Eq.~(\ref{Eq:eng-bounds-eps1}) 
will then determine a line in this two-dimensional parameter space (Fig.~\ref{Fig:bounds}a),
which we can parametrize by the parameter $V$.
The maximum possible power corresponds to $P_{\rm gen}'=0$,
where we recall $P_{\rm gen}' \equiv \rmd P_{\rm gen} \big/ \rmd V$.
This has two consequences, the first is that from Eq.~(\ref{Eq:eng-bounds-eps0}), 
we see that $P_{\rm gen}'=0$ means that  $\eps_1 \to \infty$.
Thus, the transmission function ${\cal T}_{RL}^{\mu\mu}(\eps)$,
taking the form of a Heaviside step function, $\theta(\eps-\eps_0)$,
where $\eps_0$ is given in  Eq.~(\ref{Eq:eng-bounds-eps0}).
Taking Eq.~(\ref{Eq:Pgen-eng}) combined with Eq.~(\ref{Eq:G_R})
for $\eps_1 \to \infty$,  gives
\begin{eqnarray}
P_{\rm gen}\big(\eps_1\to\infty\big)  &=& 
N\eminus V \,  \left(1-{T_R \over T_L}\right)\  G_L\left( {\eminus V \over 1-T_R/T_L}\right).
\nonumber
\end{eqnarray}
The second consequence of $P_{\rm gen}'=0$, is that the $V$-derivative of this expression 
must be zero. This gives us the condition that 
\begin{eqnarray}
(1+B_0)\ln[1+B_0] +B_0\ln[B_0] =0
\end{eqnarray}
where we define $B_0= \exp[-\eminus V/(\kB T_L-\kB T_R)] =  \exp[-\eps_0/(\kB T_L)]$.
Numerically solving this equation gives $B_0 \simeq 0.318$.
Eq.~(\ref{Eq:eng-bounds-eps0}) means that this
corresponds to $\eminus V = -\kB (T_L-T_R) \ln[0.318]= 1.146 \,\kB (T_L-T_R)$, 
indicated by the red arrow in Fig.~\ref{Fig:bounds}a.   
Substituting this back into $P_{\rm gen}\big(\eps_1\to\infty\big)$ gives
the maximum achievable value of $P_{\rm gen}$,
\begin{eqnarray}
P_{\rm gen}^{\rm qb2} = 
 A_0\, {\pi^2 \over h} N \kB^2 \big(T_L-T_R\big)^2 \quad \quad
\label{Eq:P-qb2}
\end{eqnarray}
with 
\begin{eqnarray}
A_0 \equiv B_0\ln^2[B_0]\big/\big[\pi^2(1+B_0)\big] \simeq 0.0321.
\end{eqnarray}
We refer to this as the quantum bound (qb) on power output\cite{footnote:qb2}, 
because of its origin in the Fermi wavelength of the electrons,  $\lambda_{\rm F}$.
We see this in the fact that $P_{\rm gen}^{\rm qb2}$ is proportional to 
the number of transverse modes in the quantum system, $N$,
which is given by the cross-sectional area of the quantum system divided by $\lambda_{\rm F}^2$.  This quantity has no analogue in classical thermodynamics.

The efficiency at this maximum power, $P_{\rm gen}^{\rm qb2}$, is
\begin{eqnarray}
\eta_{\rm eng} (P_{\rm gen}^{\rm qb2}) 
&=& \eta_{\rm eng}^{\rm Carnot}\big/ \big( 1+C_0 (1+T_R/T_L) \big),
\label{Eq:Eff-at-Pqb2}
\end{eqnarray}
with 
\begin{eqnarray}
C_0=-(1+B_0){\rm Li}_2(-B_0)\big/\big(B_0\ln^2[B_0]\big) \simeq 0.936.
\end{eqnarray}
As such, it varies with $T_R/T_L$, but is always more than $0.3\,\eta_{\rm eng}^{\rm Carnot}$.
This efficiency is less than Curzon and Ahlborn's efficiency for all $T_R/T_L$
(although not much less).
However, the power output here is infinitely larger than the maximum power output of systems
that achieve Curzon and Ahlborn's efficiency, see Section~\ref{Sect:eff-CA}.

The form of Eq.~(\ref{Eq:Eff-at-Pqb2}) is very different from
Curzon and Ahlborn's efficiency. 
However, we note in passing that Eq.~(\ref{Eq:Eff-at-Pqb2}) can easily be 
written as 
$\eta_{\rm eng}  (P_{\rm gen}^{\rm qb2}) 
= \eta_{\rm eng}^{\rm carnot} \big/ \left[(1+2C_0) -C_0\eta_{\rm eng}^{\rm carnot}\right]$, 
which is reminiscent of  the efficiency at maximum power found
for very different systems (certain classical stochastic heat-engines) in Eq.~(31) of 
Ref.~[\onlinecite{Schmiedl-Seifert2008}].

\subsection{Optimal heat-engine at low power output}
\label{Sect:eff-at-given-power}

Now we turn to the opposite limit, that of low power output, $P_{\rm gen}\ll P_{\rm gen}^{\rm qb2}$,
where we expect the maximum efficiency to be close to Carnot efficiency.  In this limit,
$\eps_1$ is close to $\eps_0$.
Defining $\Delta= \eps_1- \eps_0$, we 
expand Eqs.~(\ref{Eq:JL-eng},\ref{Eq:Pgen-eng}) in small $\Delta$ up to order $\Delta^3$.
This gives
\begin{eqnarray}
J_L &=& {P_{\rm gen} \over 1-T_R/T_L} 
\ +\ {N\,\Delta^3\, (1-T_R/T_L) \over 3h\,\kB T_R} g\!\left(x_0\right), \qquad
\label{Eq:JL-eng-lowpower}
\\
P_{\rm gen} &=& {N\,\eps_0 \,\Delta^2 \,(1-T_R/T_L)^2 \over 2h\,\kB T_R} 
\nonumber \\
& & \times \bigg[ g\!\left(x_0\right)  
+ {\Delta \, (1+T_R/T_L) \over 3 \, \kB T_R} \,{{\rm d} g(x_0)\over {\rm d} x_0} 
 \bigg], 
\label{Eq:Pgen-eng-lowpower}
\end{eqnarray}
where Eq.~(\ref{Eq:eng-bounds-eps0}) was used to write $\eminus V$ in terms of
$\eps_0$, 
and we defined $x_0=\eps_0/(\kB T_L)$,
and $g(x)=\e^x/(1+\e^x)^2$.
Thus, for small $\Delta$ we find that,
\begin{eqnarray}
\eta_{\rm eng}(\Delta) = \eta_{\rm eng}^{\rm Carnot} \left(1-{2\Delta \over 3x_0\kB T_L} + \cdots\right).
\label{Eq:Efficiency-in-terms-of-Delta}
\end{eqnarray}
Eq.~(\ref{Eq:eng-bounds-eps1}) gives a 
transcendental equation for $x_0$ and $\Delta$. 
However, $\Delta$ drops out when it is small, and the transcendental equation
reduces to 
\begin{eqnarray}
x_0 \tanh[x_0/2]=3,
\label{Eq:transcendental-small-Delta}
\end{eqnarray} 
for which  $x_0 \equiv \eps_0/(\kB T_L) \simeq 3.24$.   
Eq.~(\ref{Eq:eng-bounds-eps0}) means that this
corresponds to $\eminus V =3.24 \,\kB (T_L-T_R)$, indicated by the circle in Fig.~\ref{Fig:bounds}a.   
Now we can use Eq.~(\ref{Eq:Pgen-eng-lowpower}) to lowest order in $\Delta$, 
to rewrite Eq.~(\ref{Eq:Efficiency-in-terms-of-Delta}) in terms of 
$P_{\rm gen}$. This gives the efficiency for small $P_{\rm gen}$ as,
\begin{eqnarray}
\eta_{\rm eng} \big(P_{\rm gen}\big) =  \eta_{\rm eng}^{\rm Carnot} 
\left(1- 0.478
\sqrt{  {T_R \over T_L} \ {P_{\rm gen} \over P_{\rm gen}^{\rm qb2}}} \,+ \cdots 
\right)\!, \quad
\label{Eq:eta-eng-small-Pgen}
\end{eqnarray}
where the dots indicate terms of order $(P_{\rm gen} /P_{\rm gen}^{\rm qb2})$ or higher.
Eq.~(\ref{Eq:dotS-eng}) then gives the minimum rate of entropy production at power output $P_{\rm gen}$, 
\begin{eqnarray}
\dot S \big(P_{\rm gen}\big) =  0.478{P_{\rm gen}^{\rm qb2} \over \sqrt{T_LT_R} }
\left( {P_{\rm gen} \over P_{\rm gen}^{\rm qb2}}\right)^{3/2}  \,+ {\cal O}[P_{\rm gen}^2] , \quad
\label{Eq:dotS-eng-small-Pgen}
\end{eqnarray}
Thus, the maximal efficiency at small $P_{\rm gen}$
is that of Carnot minus a term that 
grows like $P_{\rm gen}^{1/2}$ (dashed curves in Fig.~\ref{Fig:allpowers}a), 
and the associated minimal rate of entropy production goes like $P_{\rm gen}^{3/2}$.

Note that Eq.~(\ref{Eq:Efficiency-in-terms-of-Delta}), 
shows that Carnot efficiency occurs at
any $x_0$ (i.e.\ any $\eps_0$) when  $\Delta$ is strictly zero (and so $P_{\rm gen}$ is strictly zero).
However, for arbitrary $x_0$ the factor 0.478 
in Eq.~(\ref{Eq:eta-eng-small-Pgen}) is replaced by 
$\sqrt{ 8\pi^2 A_0/[9 x_0^3 g(x_0)]}$.  The value of $x_0$ that satisfied 
Eq.~(\ref{Eq:transcendental-small-Delta}) 
is exactly the one which minimizes 
this prefactor (its minimum being 0.478), and thus maximizes the efficiency for any small but 
finite $P_{\rm gen}$.


\section{Guessing the optimal transmission for a refrigerator}
\label{Sect:guess-fri}

Here we use simple arguments to guess the
transmission function which maximizes a refrigerator's efficiency for given cooling power.
The arguments are similar to those for heat-engines (Section~\ref{Sect:guess-heat}), although some crucial differences will appear.

We consider the flow of electrons 
from reservoir $L$ to reservoir $R$ (the filled circle in Fig.~\ref{Fig:thermocouple}a, 
remembering $\eminus<0$ so electrons flow in the opposite direction to $I$).
To refrigerate, the thermoelectric must absorb power, so the electrical current must be due to a bias, this 
requires $\eminus V$ to be negative,
with $V$ as in Eq.~(\ref{Eq:def-V}).

Inspection of the integrand of Eq.~(\ref{Eq:JL})
shows that it only gives positive contributions to the cooling power 
output, $J_L$, when $\big(f^\mu_L(\eps) - f^\mu_R(\eps)\big) >0$.
Since $T_L< T_R$ and $\eminus V<0$, we can use Eq.~(\ref{Eq:Fermi}) to show that this is 
never true for holes ($\mu=-1$), and is only true for 
electrons ($\mu=1$) with energies $\eps < \eps_1$, where
\begin{eqnarray}
\eps_1 = -\eminus V \big/ (T_R/T_L-1).
\label{Eq:Eps1-guess}
\end{eqnarray}
Thus, it is counter-productive to allow the transmission of electrons with $\eps > \eps_1$, or the transmission of any holes.  
Note that this argument gives us an {\it upper} cut-off on electron transmission energies, despite the fact it gave a {\it lower} cut-off for the heat engine  (see Eq.~(\ref{Eq:Eps0-guess}) and the text around it).
All electron  ($\mu=1$) energies from zero to $\eps_1$ contribute 
positively to  the cooling power $J_L$.  
To maximize the cooling power, one needs to maximize $\big(f^\mu_L(\eps) - f^\mu_R(\eps)\big)$,
this is done by taking $\eminus V \to -\infty$ , for which $\eps_1 \to \infty$.
This logic gives the maximum cooling power, which Section~\ref{Sect:fri}
will show equals $\half J_L^{\rm qb}$.

Now we consider maximizing the efficiency at a given cooling output $J_L$, 
when $J_L <\half J_L^{\rm qb}$.
Comparing the integrands 
in Eqs.~(\ref{Eq:JL},\ref{Eq:Pgen}), we see that the extra factor of $\eps$ in $J_L$, means that allowing the transmission of electrons at low energies has a small effect on cooling power, while costing
a similar electrical power as higher energies. 
Thus, it would seem to be optimal to have a lower cut-off on 
transmission, $\eps_0$, which would be just low enough to ensure the 
desired cooling power $J_L$, but no lower.  
Then the transmission function will acts as a 
``band-pass filter''  (the ``box-car'' in Fig~\ref{Fig:tophat-width}),
with $\eps_0$ and $\eps_1$ further apart for higher cooling power.  This is correct, 
however the choice of $V$ affects $\eps_0$ and $\eps_1$, so the
calculation in Section~\ref{Sect:fri} is necessary to find the $V$, $\eps_0$ and $\eps_1$ which 
maximize the efficiency for cooling power $J_L$.

\section{Maximizing refrigerator efficiency for given cooling power}
\label{Sect:fri}

Here we find the maximum refrigerator efficiency,
also called the coefficient of performance (COP), for given cooling power $J_L$.
The method is very similar to that for heat-engines, and here we mainly summarize the differences. 
The refrigerator efficiency increases for a fixed cooling power, $J_L$, if the electrical 
power absorbed 
$P_{\rm abs}=-P_{\rm gen}$ decreases for fixed $J_L$. 
This is so if
 \begin{eqnarray}
 \left.{\partial P_{\rm abs} \over \partial \tau^{\mu}_\gamma }\right|_{J_L} 
&=&
 \left.{\partial P_{\rm abs} \over \partial \tau^{\mu}_\gamma }\right|_V  
- {P'_{\rm abs} \over J'_L} 
\left.{\partial J_L \over \partial \tau^{\mu}_\gamma }\right|_V \ <\ 0. \qquad
\label{Eq:fri-condition}
\end{eqnarray}
where we recall that the primed means $(\rmd / \rmd V)$.
This is nothing but Eq.~(\ref{Eq:eng-condition}) 
with $J_L \to P_{\rm abs}$ and $P_{\rm gen} \to J_L$. 
Using Eq.~(\ref{Eq:change-JL-to-change-Pgen}), we see that
$\eta_{\rm fri}(J_L)$ grows with $\tau^{\mu}_\gamma$  for
\begin{eqnarray}  
\left( {-\mu \eminus V \over \eps_\gamma}  - {P'_{\rm abs} \over J'_L} \right) \times
\left.{\partial J_L \over \partial \tau^{\mu}_\gamma }\right|_V   \ <\ 0,
\label{Eq:fri-condition2}
\end{eqnarray}
where $P_{\rm abs}$, $P'_{\rm abs}$, $J_L$, $J'_L$ and $-\eminus V$ are all positive.

To proceed we define the following energies
\begin{eqnarray}
\eps_0 &=& -\eminus V   \,J'_L / P'_{\rm abs},
\label{Eq:fri-bounds-eps0}
\\
\eps_1 &=&  {-\eminus V  \big/ (T_R/T_L-1)}.
\label{Eq:fri-bounds-eps1}
\end{eqnarray}
Then one can see that
$ \left.\left({\partial J_L/\partial \tau^{\mu}_\gamma}\right)\right|_V$ is positive
when both $\mu=1$ and $\eps < \eps_1^{\rm fri}$, 
and is negative otherwise.
Thus, for $\mu=-1$, Eq.~(\ref{Eq:fri-condition2}) is never satisfied.
For $\mu=1$,  Eq.~(\ref{Eq:fri-condition2}) is satisfied when $\eps_\gamma$ is between
$\eps_0^{\rm fri}$ and $\eps_1^{\rm fri}$.
A refrigerator is only useful if  $J_L>0$ (i.e.\ it removes heat from the cold reservoir), 
and this is only true for  $\eps_0^{\rm fri} <\eps_1^{\rm fri}$.
Hence, if $\mu=1$ and $\eps_0^{\rm fri} <\eps<\eps_1^{\rm fri}$, then 
$\eta_{\rm fri}(J_L)$ grows upon increasing  $\tau^{\mu}_\gamma$.  
Thus, the optimum is when such $\tau^{\mu}_\gamma=N$.
For all other $\mu$ and $\eps_\gamma$, $\eta_{\rm fri}(J_L)$ 
grows upon decreasing $\tau^{\mu}_\gamma$.
Thus, the optimum is when such $\tau^{\mu}_\gamma=0$.
This gives the boxcar transmission function in Eq.~(\ref{Eq:top-hat}), with
$\eps_0$ and $\eps_1$ given by Eqs.~(\ref{Eq:fri-bounds-eps0},\ref{Eq:fri-bounds-eps1}).
Comparing with Eqs.~(\ref{Eq:eng-bounds-eps0},\ref{Eq:eng-bounds-eps1}),
we see these energies are the opposite way around for a refrigerator compared to
a heat-engine (up to a minus sign).
 
Substituting Eqs.~(\ref{Eq:JL-eng},\ref{Eq:Pgen-eng})  into 
Eq.~(\ref{Eq:fri-bounds-eps0}), one gets a transcendental equation for
$\eps_0$ as a function of $V$ for given $T_R/T_L$.
This equation is too hard to solve analytically
(except in the high and low power limits, discussed in Sections \ref{Sect:qb-fri} and  \ref{Sect:lowpower-fri}).
The red curve in Fig.~\ref{Fig:bounds}b is 
a numerical solution for $T_R/T_L=1.5$. 

Having found $\eps_0$ as a function of $V$ for given $T_R/T_L$,
we can use  Eqs.~(\ref{Eq:JL-eng},\ref{Eq:Pgen-eng}) to get
$J_L(V)$ and $P_{\rm abs}(V)=-P_{\rm gen}(V)$.  We can then invert the first relation to get
$V(J_L)$. Now, we can find $P_{\rm abs}(J_L)$,
and then use Eq.~(\ref{Eq:eff-fri}) to get the quantity that we desire --- the
maximum efficiency (or COP), $\eta_{\rm fri}(J_L)$, at cooling power $J_L$.

Fig.~\ref{Fig:Delta+V}b gives the values of $\Delta=(\eps_1-\eps_0)$ 
and $\eminus V$ which result from
solving the transcendental equation numerically.
As noted, $\eps_1$ is related to $\eminus V$ by  Eq.~(\ref{Eq:fri-bounds-eps1}).
The qualitative behaviour of the resulting boxcar transmission function is sketched in Fig.~\ref{Fig:tophat-width}.
This numerical evaluation enables us to find efficiency as a function of $J_L$ and $T_R/T_L$,
which we plot in Fig.~\ref{Fig:allpowers}b.

\subsection{Quantum bound on refrigerator cooling power}
\label{Sect:qb-fri}

To find the maximum allowed cooling power, $J_L$, we look for the place where $J'_L=0$.
From Eq.~(\ref{Eq:fri-bounds-eps0}) we see that this immediately implies $\eps_0 =0$.
Taking Eq.~(\ref{Eq:JL-eng}) with $\eps_0=0$, we note by using Eq.~(\ref{Eq:Fintegral})
that $F_L(0)-F_R(0)$ grows monotonically as one takes $-\eminus V \to \infty$.
Similarly, for $\eps_1$ given by  Eq.~(\ref{Eq:eng-bounds-eps1}),
we note by using Eq.~(\ref{Eq:Fintegral}) and $T_R > T_L$ that 
$F_R(\eps_1)-F_L(\eps_1)$ grows monotonically as one takes $-\eminus V \to \infty$.
Thus, we can conclude that $J_L$ is maximal for $-\eminus V \to \infty$,
which implies $\eps_1 \to \infty$ via Eq.~(\ref{Eq:fri-bounds-eps1}).
Physically, this corresponds to all electrons arriving at the quantum system
from reservoir $L$ being transmitted into reservoir $R$, but all holes arriving from reservoir $L$ 
being reflected back into reservoir $L$. At the same time, reservoir $R$ is so strongly
biased that it has no electrons with $\eps>0$ (i.e.\ no electrons above reservoir $L$'s chemical potential) to 
carry heat from R to L. 

In this limit, $F_L(\eps_1)=F_L(\eps_1)=F_R(\eps_0)=0$,
so the maximal refrigerator cooling power is 
\begin{eqnarray}
J_L = {\pi^2 \over 12 h} N \kB^2 T_L^2 ,
\label{Eq:J-qb-fri}
\end{eqnarray}
where we used the fact that ${\rm Li}_2[1] = \pi^2/12$.
This is exactly half the quantum bound on heat current that can flow out of reservoir $L$ given in
Eq.~(\ref{Eq:Jqb}).  The quantum bound is achieved  by coupling reservoir $L$ to another reservoir with a temperature of 
absolute zero, through an contact with $N$ transverse mode.  
By definition a refrigerator is cooling reservoir $L$ below the temperature of the other reservoirs around it.  
In doing so, we show its cooling power is always less than or 
equal to $J_L^{\rm qb}/2$. However, it is intriguing that the maximum cooling power is independent
of the temperature of the environment, $T_R$, of the reservoir being cooled (reservoir $L$). 
In short, the best refrigerator can remove all electrons (or all holes) that reach it from reservoir $L$, 
but it cannot remove all electrons {\it and} all holes at the same time.

It is easy to see that the efficiency of the refrigerator (COP) at this maximum possible cooling power
is zero, simply because $|V| \to \infty$, so the power absorbed $P_{\rm abs} \to \infty$.  
However, one gets exponentially close to this limit for
$-\eminus V \gg \kB T_R$, for which $P_{\rm abs}$ is large but finite, and so $\eta_{\rm fri}(J_L)$ remains finite (see  Fig.~\ref{Fig:allpowers}b).

\subsection{Optimal refrigerator at low cooling power}
\label{Sect:lowpower-fri}

Now we turn to the opposite limit, that of low cooling power output, $J_L\ll J_L^{\rm qb}$,
where we expect the maximum efficiency to be close to Carnot efficiency.  In this limit,
$\eps_0$ is close to $\eps_1$.
Defining $\Delta= \eps_1- \eps_0$, we 
expand Eqs.~(\ref{Eq:JL-eng},\ref{Eq:Pgen-eng}) in small $\Delta$ up to order $\Delta^3$.
This gives
\begin{eqnarray}
J_L &=& {P_{\rm abs} \over  T_R/T_L-1} 
\ -\ {N\,\Delta^3 \,(T_R/T_L-1) \over 3h\,\kB T_R} g\!\left(x_1\right), \qquad
\label{Eq:JL-fri-lowpower}
\\
P_{\rm abs}&=& {N\,\eps_1 \,\Delta^2 \,(T_R/T_L-1)^2 \over 2h\,\kB T_R} 
\nonumber \\
& & \times \bigg[ g\!\left(x_1\right)  
- {\Delta \, (T_R/T_L+1) \over 3 \, \kB T_R} \,{{\rm d} g(x_1)\over {\rm d} x_1} 
 \bigg], 
\label{Eq:Pabs-fri-lowpower}
\end{eqnarray}
where Eq.~(\ref{Eq:fri-bounds-eps1}) was used to write $\eminus V$ in terms of
$\eps_1$, 
and we define $x_1=\eps_1/(\kB T_L)$,
and $g(x)=\e^x/(1+\e^x)^2$.
Thus, for small $\Delta$ we find that
the efficiency is 
\begin{eqnarray}
\eta_{\rm fri}(\Delta) = \eta_{\rm fri}^{\rm Carnot} \left(1-{2\Delta \over 3x_1\kB T_L} + \cdots\right).
\label{Eq:COP-in-terms-of-Delta}
\end{eqnarray}
Note that this is the same Eq.~(\ref{Eq:Efficiency-in-terms-of-Delta}) for the heat-engine at low power output,
except that $x_0$ is replaced by $x_1$, and the Carnot efficiency is that of the refrigerator
rather than that of the heat-engine.  

Eq.~(\ref{Eq:fri-bounds-eps0}) gives a 
transcendental equation for $x_1$ and $\Delta$.
However, $\Delta$ drops out when it is small, 
and the transcendental equation reduces to 
\begin{eqnarray}
x_1 \tanh [x_1/2]=3,
\label{Eq:condition-fri-smallJ}
\end{eqnarray}
for which $x_1\equiv \eps_1/(\kB T_L) = 3.2436\cdots$. 
Again this is the same as for a heat-engine, Eq.~(\ref{Eq:transcendental-small-Delta}), 
but with $x_1$ replacing $x_0$.
Eq.~(\ref{Eq:fri-bounds-eps1}) means that this
corresponds to $-\eminus V =3.2436 \,\kB (T_R-T_L)$, indicated by the circle in Fig.~\ref{Fig:bounds}b.   
Now we can use Eq.~(\ref{Eq:JL-fri-lowpower}) to lowest order in $\Delta$, 
to rewrite Eq.~(\ref{Eq:COP-in-terms-of-Delta}) in terms of 
$J_L$. This gives the efficiency (or coefficient of performance, COP) 
for small $J_L$ as,
\begin{eqnarray}
\eta_{\rm fri}(J_L) = \eta_{\rm fri}^{\rm Carnot} 
\left(1- 1.09
\sqrt{
\,{T_R \over T_R-T_L}\ {J_L \over J_L^{\rm qb}} }\, + \cdots \right)\!,
\nonumber \\
\label{Eq:eta-fri-smallJ}
\end{eqnarray}
where the dots indicate terms of order $(J_L/J_L^{\rm qb})$ or higher.
Eq.~(\ref{Eq:dotS-fri}) gives the minimum rate of entropy generation at cooling power output $J_L$, as
\begin{eqnarray}
\dot S \big(J_L\big) =  
1.09 
{J^{\rm qb}_L \over T_L}\sqrt{1-{T_L\over T_R}}  
\left({J_L \over J_L^{\rm qb}} \right)^{3/2}\, + {\cal O}[J_L^2],
\nonumber \\
\label{Eq:dotS-fri-smallJ}
\end{eqnarray}
Thus, we conclude that the maximum efficiency at small $J_L$ is that of Carnot minus a term that 
grows like $J_L^{1/2}$  (dashed curves in Fig.~\ref{Fig:allpowers}b), while the associated
minimum entropy production goes like $J_L^{3/2}$.

We note that Carnot efficiency occurs at $J_L =0$ at any $x_1=\eps_1/(\kB T_L)$.
However, then the 1.09 factor in Eq.~(\ref{Eq:eta-fri-smallJ})
becomes $\sqrt{ 4\pi^2/[27 x_1^3 g(x_1)]}$.
The condition in Eq.~(\ref{Eq:condition-fri-smallJ}) minimizes this factor (the minimum being 1.09),
and thereby maximizes the efficiency for given  $J_L$.


\section{Implementation with a chain of quantum systems}
\label{Sect:chain}

\begin{figure}[t]
\includegraphics[width=\columnwidth]{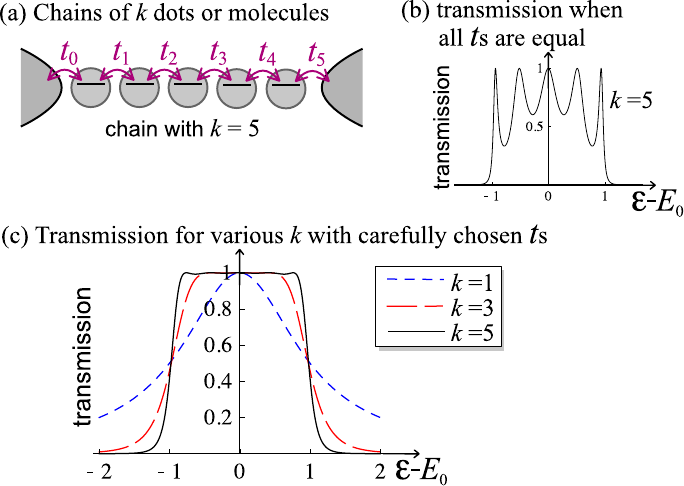}
\caption{\label{Fig:band}
(a) A chain of single level quantum dots with their energy levels aligned at energy $E_0$. 
(b) Transmission function when all hoppings are equal (note the strong oscillations).
(c)  Transmission function when all hoppings are carefully chosen (see text).
To aid comparison all bandwidths in the plots have been normalized.
}
\end{figure}

The previous sections have shown that maximum efficiency
(at given power  output) occurs when the thermoelectric system has a boxcar transmission function
with the right position and width.  In the limit of maximum power, the boxcar becomes a  
Heaviside step-function. Here, we give a detailed recipe for engineering such transmission functions
for non-interacting electrons, and then discuss how to include mean-field interaction effects.

A Heaviside step-function is easily 
implemented with point-contact, whose transmission function is
\cite{Buttiker-pointcont},
\begin{eqnarray}
{\cal T}_{\rm L,isl}(\eps) = \left(1+ \exp \left[- {\eps-E(V) \over D_{\rm tunnel} }\right] \right)^{-1} 
\label{Eq:transmission-pc}
\end{eqnarray}
where $E(V)$ is the height of the energy barrier induced by the point contact,
and $D_{\rm tunnel}$ is a measure of tunnelling through the point contact.
A sufficiently long point contact exhibits negligible tunnelling, $D_{\rm tunnel}  \to 0$, 
so the transmission
function simplifies to the desired Heaviside step-function, $\theta[\eps-E(V)]$.

For a potential implementation of a boxcar function we consider a
chain of sites (quantum dots or molecules) with one level per site, as sketched in Fig.~\ref{Fig:band}a.
The objective is that 
the hoppings between sites, $\{t_i\}$, 
will cause the states to hybridize to form a band centred at $E_0$, with a width 
given by the hopping\cite{Buttiker-private-comm}.
Neglecting electron-electron interactions, 
the hopping Hamiltonian for five sites in the chain ($k=5$) can be written as
\begin{eqnarray}
{\cal H}_{\rm chain} = \left(\begin{array}{ccccc}
-\rmi a_0 /2 \ & t_1 & 0 & 0 & 0 \\
t_1 & 0 & t_2 & 0 & 0 \\
0 &  t_2 & 0 & t_3 & 0 \\
0 & 0 &   t_3 & 0 & t_4  \\
0 & 0 & 0 & t_4 &  \ -\rmi a_0/2
\end{array}\right).
\end{eqnarray}
This is easily generalized to arbitrary chain length, $k$.
Here we treat $a_0$ as a phenomenological parameter, however in reality it would be given by $|t_0|^2$ multiplied by the density of states in the reservoir.
The fact that particles escape from the chain into the reservoirs, means the wavefunction for any given particle 
in the chain will decay with time.
To model this, the Hamiltonian must be non-Hermitian, with the non-Hermiticity entering in the matrix elements for coupling to the reservoirs (top-left and bottom right matrix elements).
These induce an imaginary contribution to each eigenstate's energy $E_i$, with  
the wavefunction of any eigenstate decaying at a rate given by the imaginary part of $E_i$.
The non-Hermiticity of $ {\cal H}_{\rm chain}$ also means that its left and right eigenvectors are different,
defining  $\big | \psi_i^{\rm (r)}\big \rangle$ as the $i$th right eigenvector of the matrix
${\cal H}_{\rm chain}$, and $\big\langle\psi_i^{\rm (l)} \big|$ as the $i$th  left eigenvector,
we have
$\big\langle\psi_i^{\rm (l)}  \big | \psi_j^{\rm (r)}\big \rangle = \delta_{ij}$ and 
$\big\langle\psi_i^{\rm (l)}  \big | {\cal H}_{\rm chain}  \big | \psi_i^{\rm (r)}\big \rangle = E_i$.
The resolution of unity is 
$\sum_{i}  \big | \psi_i^{\rm (r)}\big \rangle \, \big\langle\psi_i^{\rm (l)}  \big | = {\bm 1}$, 
where ${\bm 1}$ is the $k$-by-$k$ unit matrix.

We define $|1\rangle$ as the vector whose first element is one while all its other elements are zero, and 
$|k\rangle$ as the vector whose last element (the $k$th element) is one while all its other elements are zero. 
Then the transmission probability at energy $\eps$ is given by 
\begin{eqnarray}
{\cal T}_{RL}(\eps) &=& 
\left| \big\langle k \big|  \ \left[\eps -{\cal H}_{\rm chain}\right]^{-1} \big| 1 \big\rangle 
\right|^2 \ a_0 \, ,
\end{eqnarray}
where $[\cdots]^{-1}$ is a matrix inverse.
To evaluate this matrix inverse, we introduce a resolution of unity
to the left and right of $\left[\eps -{\cal H}_{\rm chain}\right]^{-1}$.
This gives 
\begin{eqnarray}
{\cal T}_{RL} &=& 
\sum_i \
\left|{\big\langle k \big | \psi_i^{\rm (r)}\big\rangle \ \big\langle\psi_i^{\rm (l)} \big| 1 \big\rangle 
\over \eps-E_i} \right|^2 \ a_0.
\label{Eq:T-for-chain}
\end{eqnarray}
For any given set of hoppings $a_0, t_1,\cdots t_k$, one can easily 
use a suitable eigenvector finder (we used Mathematica) to evaluate this equation numerically,
while an analytic solution is straight-forward\cite{Grenier-private} for $k\leq 3$. 
When all hoppings in the chain are equal, 
there is a mismatch between the electron's hopping dynamics in the chain and their free motion in the reservoirs.   This causes resonances in the transmission, giving the  
Fabry-Perot-type oscillations in Fig.~\ref{Fig:band}b for $k=5$. 
However, we can carefully tune the hoppings (to be smallest in the middle of the chain and increasing towards the ends) to get the smooth transmission functions in  Fig.~\ref{Fig:band}c.
The $k=5$ curve in Fig.~\ref{Fig:band}c has $t_1=t_4= 0.39a_0$ and $t_2=t_3= 0.28 a_0$, 
and we choose $a_0= 1.91$ to normalize the band width to 1.
As the number of sites in the chain, $k$, increases, the transmission function tends to the desired boxcar function.

The above logic assumes no electron-electron interactions.
When we include interaction effects at the mean-field level, things get more complicated.
If the states in the chain are all at the same energy $E_0$ when the chain is unbiased,
they will not be aligned when there is a bias between the the reservoirs, because the reservoirs
also act as gates on the chain states.  To engineer a chain where the energies are aligned at the optimal bias, one must adjust the confinement
potential of the dots in the chain (or adjust the chemistry of the molecules in the chain)
so that their energies are sufficiently out of alignment at zero bias that they all align 
at optimal bias.  In principle, we have the control to do this. However, in practice it would
require a great deal of trial-and-error experimental fine tuning.  
We do not enter further into such practical issues here.
Rather, we use the above example to show that there is no {\it fundamental} reason that the bound
on efficiency cannot be achieved.


\section{Many quantum systems in parallel}
\label{Sect:in-parallel}

To increase the efficiency at given power output, one must increase the number of transverse modes, $N$.
This is because the efficiency decays with the power output divided by
the quantum bounds in Eqs.~(\ref{Eq:P-qb2},\ref{Eq:J-qb-fri}), 
and these bounds go like $N$. 
However, a strong thermoelectric response requires a
transmission function that is highly energy dependent, this typically only occurs when the 
quantum system (point-contact, quantum dot or molecule) has dimensions of about a wavelength,
which implies that $N$ is of order one.
Crucial exceptions (beyond the scope of this work) are systems containing superconductors,
either SNS structures\cite{Pekola-reviews} or Andreev interferometers \cite{Chandra98} 
(see also Ref.~[\onlinecite{jw-epl}] and references therein),
where strong thermoelectric effects occur for large $N$.

In the absence of a superconductor, the only way to get large $N$ 
is to construct a device consisting of many $N=1$ systems in parallel, such as
a surface covered with a certain density of such systems 
\cite{Jordan-Sothmann-Sanchez-Buttiker2013,Sothmann-Sanchez-Jordan-Buttiker2013}.  In this case $P_{\rm gen}^{\rm qb2}$ and $J_{\rm L}^{\rm qb}$ in
Eqs.~(\ref{Eq:P-qb2},\ref{Eq:J-qb-fri}) become
bounds on the power per unit area, with $N$ being replaced by the number of transverse modes 
per unit area.
With this one modification, all calculations and results in this article
can be applied directly to such a situation. 
Carnot efficiency is achieved for a large enough surface area that
the power per unit area is much less than $P_{\rm gen}^{\rm qb2}$ and $J_{\rm L}^{\rm qb}$.

It is worth noting that the number of modes per unit area cannot exceed $\lambda_{\rm F}^{-2}$, for Fermi wavelength 
$\lambda_{\rm F}$. 
From this we can get a feeling for the magnitude of the bounds discussed in this article.
Take a typical semiconductor thermoelectric (with $\lambda_{\rm F}\sim 10^{-8}$m),
placed between reservoirs at 700 K and 300 K (typical temperatures for a thermoelectric recovering electricity
from the heat in the exhaust gases of a diesel engined car).
 Eq.~(\ref{Eq:P-qb2}) tells us that to get 100 W of power output from a semiconductor thermoelectric 
 one needs a cross section of at least  4 mm$^2$.  
Then Eq.~(\ref{Eq:eta-eng-small-Pgen}) tells us that to get this power at 90\% of Carnot efficiency, 
one needs a cross section of  at least  0.4 cm$^2$.
Remarkably, it is {\it quantum mechanics} which gives these bounds, 
even though the cross sections in question are macroscopic.


\section{Phonons and photons carrying heat in parallel with electrons}
\label{Sect:ph}

\begin{figure}
\includegraphics[width=\columnwidth]{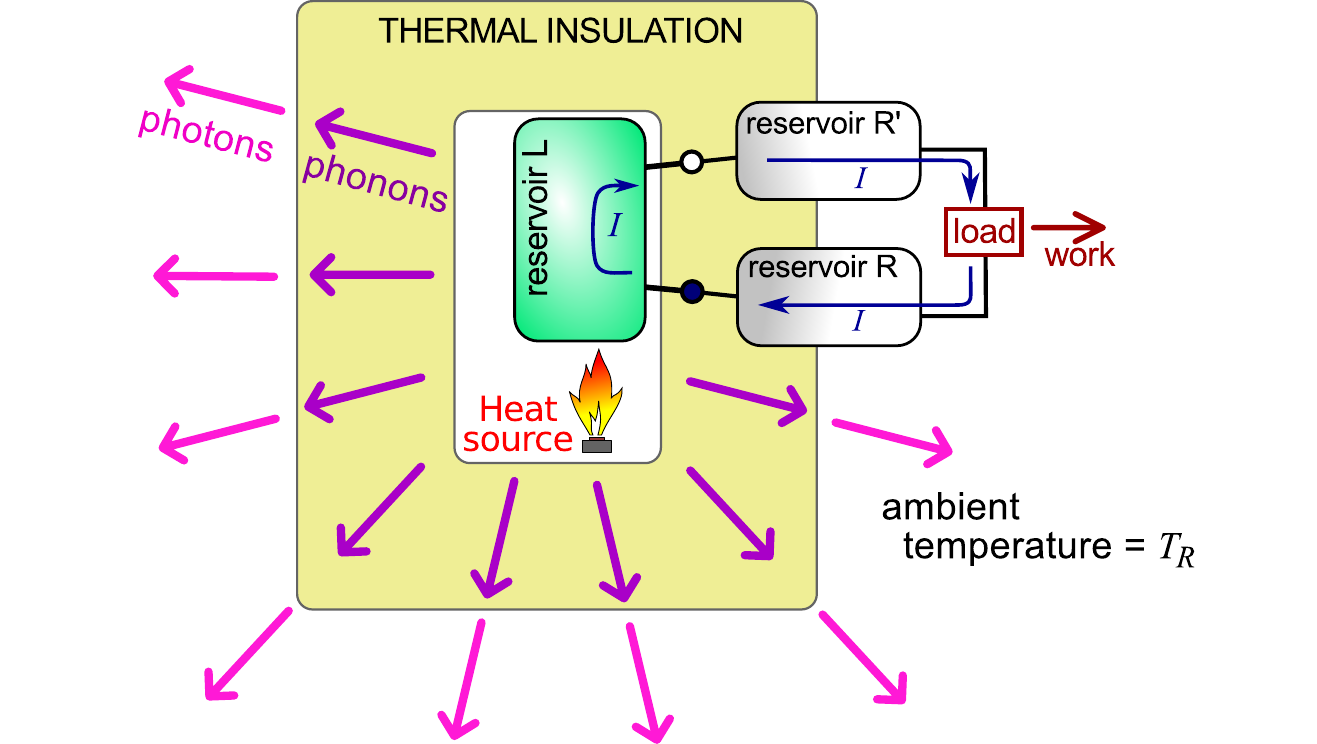}
\caption{\label{Fig:thermocouple-phonons}
The thermocouple heat-engine in Fig.~\ref{Fig:thermocouple}, showing the heat flow due 
to phonon and photons, which carry heat from hot to cold by all possible routes (in parallel with the heat carried by the electrons).  
This always reduces the efficiency, so it should be minimized with suitable thermal insulation.}
\end{figure}

Any charge-less excitation (such as phonons or photons)
will carry heat from hot to cold, irrespective of the thermoelectric properties of the system.
While some of the phonons and photons will flow through the thermoelectric quantum system, 
most will flow via other routes, see Fig.~\ref{Fig:thermocouple-phonons}.
A number of theories for these phonon or photon heat currents 
take the form
\begin{eqnarray}
J_{\rm ph}=  \alpha (T_L^\kappa-T_R^\kappa),
\label{Eq:J_ph}
\end{eqnarray}
where $J_{\rm ph}$ is  the heat flow out of the L reservoir due to phonons or photons.
The textbook example of such a theory is that of black-body radiation between the two reservoirs,
then $\kappa=4$ and $\alpha$ is the Stefan-Boltzmann constant.
An example relevant to suspended sub-Kelvin nanostructures
is a situation where a finite number $N_{\rm ph}$ of
phonon or photon modes carry heat between the two reservoirs
 \cite{Pendry1983,photons,phonons,2012w-pointcont} 
then $\kappa=2$ and $\alpha \leq N_{\rm ph}\pi^2 \kB^2/(6h)$.

One of the biggest practical challenges for quantum thermoelectrics is that phonons and photons
will often carry much more heat than the electrons.  This is simply because the hot reservoir
can typically radiate heat in all directions as phonons or photons, while electrons only carry heat 
through the few nanostructures connected to that reservoir.
Thus, in many cases the phonon or photon heat flow will dominate over the electronic one.
However, progress is being made in blocking phonon and photon flow, by suspending the nanostructure 
to minimize phonon flow \cite{phonons} and engineering the electromagnetic environment to minimize photon flow \cite{photons}, and it can be hoped that phonon and phonon effects
will be greatly reduced in the future.
Hence, here we consider the full range from weak to strong phonon or photon heat flows. 

For compactness in what follows we will only refer to phonon heat flows (usually the dominant parasitic effect).
However, strictly one should consider $J_{\rm ph}$ as the sum of the heat flow carried by phonons, photons and any more exotic charge-less excitations that might exist in a given circuit (mechanical oscillations, spin-waves, etc.).

\begin{figure}
\includegraphics[width=\columnwidth]{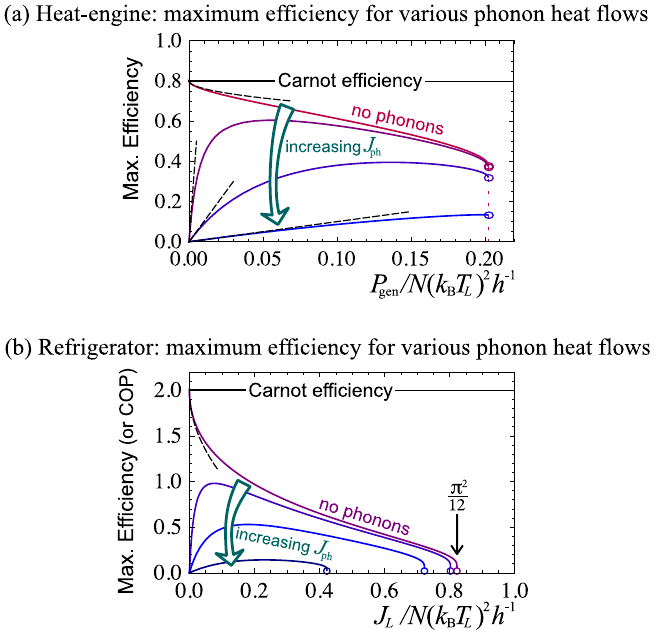}
\caption{\label{Fig:allpowers-phonons} 
Plots of the maximum efficiency allowed when the there is a phonon heat flow, $J_{\rm ph}$,
in parallel with the heat carried by the electrons.  
The curves in (a) are for $T_R/T_L=0.2$, with $J_{\rm ph} = 0, 0.01, 0.1, 1$  (from top to bottom); the curves
come from Eq.~(\ref{Eq:eng-e+ph}) with $\eta_{\rm eng}(P_{\rm gen})$ given 
in Fig.~\ref{Fig:allpowers}a.
The curves in (b) are for $T_R/T_L=1.5$, with $J_{\rm ph} = 0, 0.02, 0.1, 0.4$ (from top to bottom); the curves
come from Eq.~(\ref{Eq:fri-e+ph}) with $\eta_{\rm eng}(P_{\rm gen})$ given 
in Fig.~\ref{Fig:allpowers}b.
The maximum cooling power (open circles) is $(\half J^{\rm qb}_L-J_{\rm ph})$.
}
\end{figure}

\subsection{Heat-engine with phonons}
For heat-engines, the phonon heat-flow is in parallel with electronic heat-flow, so 
the heat-flow for a given $P_{\rm gen}$ is $(J_L+J_{\rm ph})$, rather than just $J_L$
(as it was in the absence of phonons).
Thus,  the efficiency in the presence of the phonons  is
\begin{eqnarray}
\eta^{\rm e+ph}_{\rm eng}(P_{\rm gen})={P_{\rm gen} \over J_L(P_{\rm gen})+J_{\rm ph}}.
\end{eqnarray}
Writing this in terms of the efficiency, we get
\begin{eqnarray}
\eta_{\rm eng}^{\rm e+ph} (P_{\rm gen}) 
&=& \big[ \eta_{\rm eng}^{-1} (P_{\rm gen}) +J_{\rm ph} / P_{\rm gen} \big]^{-1},  
\label{Eq:eng-e+ph}
\end{eqnarray}
where $\eta_{\rm eng} (P_{\rm gen})$ is the efficiency for $J_{\rm ph}=0$.   
Given the maximum efficiency at given power in the absence of phonons,
we can use this result to find the maximum efficiency for a given phonon heat flow, 
$J_{\rm ph}$.
An example of this is shown in Fig.~\ref{Fig:allpowers-phonons}a.
It shows that for finite $J_{\rm ph}$, Carnot efficiency is not possible at any power output.

Phonons have a huge effect on the efficiency at small power output.
Whenever $J_{\rm ph}$ is non-zero, the efficiency vanishes at zero power output,
with
\begin{eqnarray}
\eta^{\rm e+ph}_{\rm eng}(P_{\rm gen})=P_{\rm gen}\big/J_{\rm ph} \ \ \  \hbox{ for } \ P_{\rm gen} \ll J_{\rm ph}.
\label{Eq:eta-with-phonons-smallP}
\end{eqnarray}
As $J_{\rm ph}$ increases,  the range of applicability of this small $P_{\rm gen}$ approximation (shown as dashed lines in Fig.~\ref{Fig:allpowers-phonons}) grows towards the maximum power $P_{\rm eng}^{\rm qb}$ (open circles).  
In contrast, phonon heat flows have little effect on the efficiency near the maximum power output, until these flows become strong enough that $J_{\rm ph} \sim P_{\rm gen}$.
 
For strong phonon flow, where $J_{\rm ph} \gg  P_{\rm gen}$, 
Eq.~(\ref{Eq:eta-with-phonons-smallP}) applies at all powers up to the maximum, $P_{\rm gen}^{\rm qb2}$.
Then, the efficiency is maximal when the power is maximal, where maximal power is the quantum bound given in Eq.~(\ref{Eq:P-qb2}).
Thus, the system with both maximal power and maximal efficiency is that with 
a Heaviside step transmission function (see section~\ref{Sect:chain}).

\subsection{Refrigerator with phonons}

For a refrigerator to extract heat from a reservoir at rate $J$ in 
the presence of phonons carrying a back flow of heat $J_{\rm ph}$, 
that refrigerator must extract heat at a rate $J_L=J+J_{\rm ph}$.
Note that for clarity, in this section we take $J_{\rm ph}$ to be positive when $T_L< T_R$
(opposite sign of that in Eq.~(\ref{Eq:J_ph})).
Thus, the efficiency, or COP, in the presence of phonons,
is  the heat current extracted, $J$, divided by the electrical power 
required to extract heat at the rate  $J_L=(J+J_{\rm ph})$.  
This means that
\begin{eqnarray}
\eta_{\rm fri}^{\rm e+ph}(J) &=&  {J \, \eta_{\rm fri}(J+J_{\rm ph}) \over J+J_{\rm ph}} ,
\label{Eq:fri-e+ph}
\end{eqnarray}
where $\eta_{\rm fri} (J)$ is the efficiency for $J_{\rm ph}=0$.   
We can use this result to find the maximum efficiency for a given phonon heat flow, 
$J_{\rm ph}$.
An example is shown in Fig.~\ref{Fig:allpowers-phonons}b.  

Eq.~(\ref{Eq:fri-e+ph}) means that the phonon flow suppresses the maximum cooling power, 
so $J$ must now obey
\begin{eqnarray}
J&\leq& \half J_L^{\rm qb} -J_{\rm ph}
\label{Eq:Jqb-phonons}
\end{eqnarray}
with $J_L^{\rm qb}$ given in Eq.~(\ref{Eq:Jqb}).
Thus, the upper bound (open circles) in Fig.~\ref{Fig:allpowers-phonons}b 
move to the left as $J_{\rm ph}$ increases.

When the reservoir being refrigerated (reservoir $L$) is at ambient temperature, $T_R$,
then  $J_{\rm ph}=0$ while $J_L^{\rm qb}$ is finite.  However, as reservoir $L$ is refrigerated (reducing $T_L$),
$J_{\rm ph}$ grows, while $J_L^{\rm qb}$ shrinks.  
As a result, at some point (before $T_L$ gets to zero) one arrives at   $J_{\rm ph} =  \half J_L^{\rm qb}$,
and further cooling of reservoir $L$ is impossible.  Thus, given the $T_L$ of $J_{\rm ph}$ for a given system,  one can easily find the lowest temperature  
that reservoir $L$ can be refrigerated to, by solving the equation 
$J_{\rm ph} =  \half J_L^{\rm qb}$ for $T_L$
To achieve this temperature, one needs 
the refrigerator with the maximum cooling power (rather than the most efficient one), this 
is a system with a Heaviside step transmission function (see section~\ref{Sect:chain}). Such a system's refrigeration capacities were discussed in Ref. [\onlinecite{2012w-pointcont}].

We also note that, as with the heat-engine, phonons have a huge effect on the efficiency at small cooling power, 
as can be seen in Fig.~\ref{Fig:allpowers-phonons}b.
Whenever $0< J_{\rm ph}<\half J_L^{\rm qb}$, the efficiency vanishes for small cooling power, with
\begin{eqnarray}
\eta^{\rm e+ph}_{\rm fri}(J)=J \ {\eta_{\rm fri}(J_{\rm ph}) \over J_{\rm ph}}\ \ \ \ \hbox{ for } \ J \ll J_{\rm ph}.
\end{eqnarray}


\section{Relaxation in a quantum system without B-field}
\label{Sect:Relax}

Elsewhere in this article, we neglected relaxation in the quantum system. In other words, we assumed 
that electrons traverse the system in a time much less than the time for
inelastic scattering from phonons, photons or other electrons.
We now consider systems in which there is such relaxation, and ask 
if this relaxation could enable a system to exceed the bounds found above for relaxationless systems.  
To make progress, we restrict our interest to systems with negligible 
external magnetic field (B-field) \cite{Footnote:Error-my-PRL}. 
As yet, we have not been able to consider the rich interplay of relaxation and B-field
\cite{Casati2011,Sanchez-Serra2011,Entin-Wohlman2012}.

\begin{figure}
\includegraphics[width=\columnwidth]{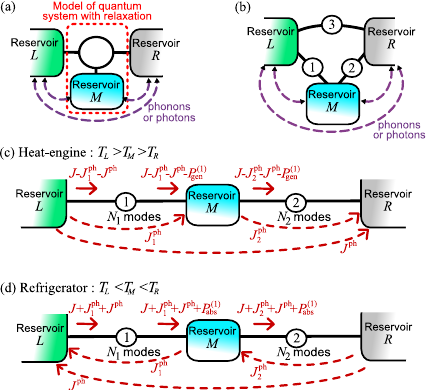}
\caption{\label{Fig:relax} 
(a) A quantum system in which relaxation occurs is modelled phenomenologically by 
a coherent quantum system coupled to a third fictitious reservoir $M$ in which the relaxation occurs.
(b) The same model after we have separated the system's scattering matrix
into three components.  The dashed arrows are the exchange of phonons or photons.
The arm containing scatterers 1 and 2 is shown  in (c) for a heat-engine,
and in (d) for a refrigerator.
}
\end{figure}

We use the voltage-probe model \cite{voltage-probe} shown in Fig.~\ref{Fig:relax}a. 
A system with relaxation is modeled as a phase-coherent scatterer coupled to a
fictitious reservoir $M$ (a region in which relaxation occurs instantaneously).  
The rate of the relaxation is controlled by the transmission
of the lead coupling to reservoir $M$.
We then separate the phase-coherent scatterer 
into scatterers 1,2 and 3, as shown in Fig.~\ref{Fig:relax}b, each with their own transmission functions
${\cal T}_{ij}(\eps)$ with $i,j \in L,M,R$.
We assume that the transmission is unchanged under reversal of direction,
so ${\cal T}_{ij}(\eps)={\cal T}_{ji}(\eps)$ for all $\eps$ and  $i,j$.
This condition is guaranteed by time-reversal symmetry whenever the B-field has a negligible effect on the electron and hole dynamics.  However, it also applies for any B-field when all particles relax as they traverse the quantum system (then ${\cal T}_{LR}(\eps)={\cal T}_{RL}(\eps)=0$, which is sufficient
to force ${\cal T}_{ij}(\eps)={\cal T}_{ji}(\eps)$ for all $i,j$).

If the relaxation involves electron-phonon or electron-photon interactions
(typically any system which is not sub-Kelvin), 
the phonons or photons with which the electrons interact usually
flow easily between the system and the reservoirs.  
Thus, these phonons or photons can carry heat current 
between the fictitious reservoir $M$ and reservoirs $L,R$ (dashed arrows in Fig.~\ref{Fig:relax}).
The total electrical and heat currents into reservoir $M$ must be zero,
and this constraint determines reservoir $M$'s bias, $V_M$, and temperature, $T_M$.

\subsection{Method of over-estimation}

The optimal choice of ${\cal T}_{ML}$ and  ${\cal T}_{RM}$ depends on
$T_M$, while $T_M$ depend on the heat current,
and thus on ${\cal T}_{ML}$ and  ${\cal T}_{RM}$.
The solution and optimization of this self-consistency problem has been beyond our ability
to resolve, even though we have restricted ourselves to a simple model of relaxation in a system with negligible B-field.
Instead, we make a simplification
which leads to an {\it over-estimate} of the efficiency.  
We assume $V_M,T_M$ are free parameters (not determined from ${\cal T}_{ML}$ and  ${\cal T}_{RM}$), 
with $T_M$ between $T_L$ and $T_R$.  
If we find the optimal  ${\cal T}_{ML}$ and  ${\cal T}_{RM}$ for given $T_M$, and 
then find the optimal $T_M$ (irrespective of whether it is consistent with ${\cal T}_{ML}$ and ${\cal T}_{RM}$ or not), we have an over-estimate of the
maximal efficiency.  
Even with this simplification, we have only been able to address the low-power and high-power limits. 
However, we show below that this over-estimate is sufficient to prove the following.
\begin{itemize}
\item[(1)] At low power, relaxation cannot make the system's efficiency exceed that of the optimal relaxation-free system
with $N_{\rm max}$ modes.
\item[(2)]  Relaxation cannot make a system's power exceed that of the maximum possible power of a 
relaxation-free system with $N_{\rm max}$ modes.
\end{itemize}
Defining $N_L$  and $N_R$ as 
the number of transverse modes in the system to the left and right of the region where relaxation occurs,
\begin{eqnarray}
N_{\rm max}={\rm max}[N_L,N_R],
\label{Eq:Nmax}
\end{eqnarray} 

\subsection{Efficiency of heat-engine with relaxation}
\label{Sect:eng-eff-relax}
To get the efficiency for our model of a quantum system with relaxation, we must find the efficiency for
the system in  Fig.~\ref{Fig:relax}b.  This system has two ``arms''.
One arm contains scatterers 1 and 2, and we define its efficiency as  $\eta_{\rm eng}^{(1\&2)}$.
The other arm contains scatterer 3,
 and we define its efficiency as  $\eta_{\rm eng}^{(3)}$.
The efficiency of the full system, $ \eta_{\rm eng}^{\rm total}(P_{\rm gen})$, is given by 
\begin{eqnarray}
{1 \over \eta_{\rm eng}^{\rm total}(P_{\rm gen}) } = 
{p_{\rm rel} \over \eta_{\rm eng}^{(1\&2)} (p_{\rm rel}P_{\rm gen}) } 
+{q_{\rm rel} \over \eta_{\rm eng}^{(3)} ( q_{\rm rel}P_{\rm gen}) }, 
\label{Eq:heatengines-in-parallel}
\end{eqnarray}
Here $p_{\rm rel}$ is the proportion of transmitted electrons that have passed through the arm containing scatterers 1 and 2, while  $q_{\rm rel}=(1-p_{\rm rel})$ is the proportion that have passed through the arm containing scatterer 3.
Physically, $p_{\rm rel}$ is 
the probability that an electron entering the quantum system relaxes before transmitting,
while $q_{\rm rel}$ is the probability that it transmits before relaxing.
One sees from Eq.~(\ref{Eq:heatengines-in-parallel}) that 
the maximal efficiency for a given $p_{\rm rel}$ occurs when both $\eta_{\rm eng}^{(1\&2)}$
and $\eta_{\rm eng}^{(3)}$ are maximal.  

The upper-bound on $\eta_{\rm eng}^{(3)}$
is that given in section~\ref{Sect:eng} with $q_{\rm rel}N_L$
modes to the left and $q_{\rm rel}N_R$ modes to the right.  
Our objective now is to find the maximum  $\eta_{\rm eng}^{(1\&2)}$
with $N_1=p_{\rm rel}N_L$ modes on the left and $N_2=p_{\rm rel}N_R$ modes on the right.
More precisely our objective is to find an {\it over-estimate} of this maximum.
For the heat flows indicated in Fig.~\ref{Fig:relax}c, the efficiency is
\begin{eqnarray}
\eta_{\rm eng}^{(1\&2)} &\equiv& P_{\rm gen}^{(1\&2)}\big/J
\nonumber \\
&=&\!\! {1 \over J}\left[
P_{\rm gen}^{(1)}(J_1;T_M,T_L) + P_{\rm gen}^{(2)}(J_2;T_R,T_M)\right]\! , \qquad \ 
\end{eqnarray}
where $J_1=J-J_1^{\rm ph}-J^{\rm ph}$ and $J_2 =J-J_2^{\rm ph}-J^{\rm ph}-P_{\rm gen}^{(1)}$.
One sees that $\eta_{\rm eng}^{(1\&2)}$  is maximal for  given $T_M$ when 
$J^{\rm ph}=J_1^{\rm ph}=J_2^{\rm ph}=0$ (these heat currents cannot be negative
because $T_L > T_M> T_R$).
Thus, to get our over-estimate of the maximal efficiency for given $T_M$,
we assume these phonon and photon heat-currents are zero. 
Then, with a little algebra, one finds that
\begin{eqnarray}
1-\eta_{\rm eng}^{(1\&2)}\big(P_{\rm gen}^{(1\&2)}\big) = \left(1-\eta_{\rm eng}^{(1)}\big(P_{\rm gen}^{(1)}\big) \right)\left(1-\eta_{\rm eng}^{(2)}\big(P_{\rm gen}^{(2)}\big) \right),
\nonumber
\end{eqnarray}
where $P_{\rm gen}^{(1)}$ and  $P_{\rm gen}^{(1)}$ are related to $P_{\rm gen}^{(1\&2)}$ by
\begin{eqnarray}
P_{\rm gen}^{(\mu)} = P_{\rm gen}^{(1\&2)} \eta_{\rm eng}^{(\mu)} \big/ \eta_{\rm eng}^{(1\& 2)},
\label{Eq:P1-or-2}
\end{eqnarray}
for $\mu=1,2$.
For given $T_M$, one maximizes $\eta_{\rm eng}^{(1\&2)}$ by independently maximizing
 $\eta_{\rm eng}^{(1)}$ and  $\eta_{\rm eng}^{(2)}$. 
For low powers, Eq.~(\ref{Eq:eta-eng-small-Pgen}) with $P,N,T_R\to P_1, N_1, T_M$ gives $\eta_{\rm eng}^{(1)}$,
while with $P,N,T_L\to P_2, N_2, T_M$ gives $\eta_{\rm eng}^{(2)}$. 
In this limit, we can 
treat efficiencies in Eq.~(\ref{Eq:P1-or-2}) to zeroth order in  $P_{\rm gen}^{(1\&2)}$, taking 
them to be Carnot efficiencies, so 
\begin{eqnarray}
P_{\rm gen}^{(1)} \simeq {T_L-T_M \over T_L-T_R}P_{\rm gen}^{(1\&2)},  \quad
P_{\rm gen}^{(1)} \simeq {T_M-T_R \over T_L-T_R}P_{\rm gen}^{(1\&2)}. 
\nonumber
\end{eqnarray}
Then some algebra gives the over-estimate of efficiency at low powers for given $T_M$, to be
\begin{eqnarray}
\eta_{\rm eng}^{(1\&2)} \leq
\eta_{\rm eng}^{\rm Carnot} 
\left(1- 0.478
\sqrt{  {T_R \over T_L} \ {P_{\rm gen} \ K_{\rm rel}\over P_{\rm gen}^{\rm qb2}(N=1)} } \right)\! ,
\quad
\end{eqnarray}
with $P_{\rm gen}^{\rm qb2}(N=1)$ given by Eq.~(\ref{Eq:P-qb2}) with $N=1$, and
\begin{eqnarray}
K_{\rm rel} = 
\sqrt{{1 \over N_1}\,{T_R(T_L-T_M) \over T_M(T_L-T_R)}} 
+ \sqrt{{1 \over N_2}\,{T_L(T_M-T_R) \over T_M(T_L-T_R)}},\quad 
\label{Eq:Krelax}
\end{eqnarray}
where $N_1= p_{\rm rel} N_L$ and $N_2=p_{\rm rel} N_L$ are respectively the number of transmission modes in scattering matrices 1 and 2. 
The over-estimate of $\eta_{\rm eng}^{(1\&2)}$ is maximal when $T_M$ is chosen to minimize 
$K_{\rm rel}$.
The two minima of  $K_{\rm rel}$ are at $T_M=T_R$ and $T_M=T_L$, 
for which the values of $K_{\rm rel}$ 
are $1/\sqrt{N_1}$ and  $1/\sqrt{N_2}$ respectively. 
Thus, we have 
\begin{eqnarray}
K_{\rm rel} \geq 1/\sqrt{p_{\rm rel}N_{\rm max}}\ ,
\label{Eq:Krelax-limit} 
\end{eqnarray}
with $N_{\rm max}$ in Eq.~(\ref{Eq:Nmax}).  Thus, whatever $T_M$ may be, 
\begin{eqnarray}
\eta_{\rm eng}^{(1\&2)} \left(P_{\rm gen}^{(1\&2)}\right) &\leq&
\eta_{\rm eng}^{\rm Carnot}
\nonumber \\
& & \times 
\left(\! 1- 0.478
\sqrt{  {T_R \over T_L}  {P_{\rm gen}^{(1\&2)} \over P_{\rm gen}^{\rm qb2}(p_{\rm rel}N_{\rm max}) } }  \right)\! .
\nonumber \\
\label{Eq:eta1&2-bound}
\end{eqnarray}
Since $P_{\rm gen}^{(1\&2)} = p_{\rm rel} P_{\rm gen}$, we can simplify Eq.~(\ref{Eq:eta1&2-bound}) 
by noting that 
\begin{eqnarray}
{P_{\rm gen}^{(1\&2)} \over P_{\rm gen}^{\rm qb2}(p_{\rm rel}N_{\rm max}) } = 
{P_{\rm gen} \over P_{\rm gen}^{\rm qb2}(N_{\rm max}) } 
\end{eqnarray}
where $P_{\rm gen}$ is the total power generated by the combined system made of scatterers 1,2 and 3. 
Then substituting the result into Eq.~(\ref{Eq:heatengines-in-parallel}), we get an over-estimate of the 
efficiency at power output $P_{\rm gen}$ which is equal to the upper bound we found in the absence of relaxation, Eq.~(\ref{Eq:eta-eng-small-Pgen}).

Thus, we can conclude that for small power outputs,
no quantum system with relaxation within it can exceed the upper-bound on efficiency found for a {\it relaxation-free} system with $N_{\rm max}$ transverse modes.
Since the proof is based on an over-estimate of the efficiency for a system with relaxation,
we cannot say if a system with finite relaxation can approach the bound in Eq.~(\ref{Eq:eta-eng-small-Pgen}).
Unlike in the relaxation-free case, we cannot say what properties the quantum system with relaxation 
(as given in terms of the properties of the effective scatterers 1, 2 and 3) are necessary to maximize the efficiency
at given power output.  We simply know that it cannot exceed Eq.~(\ref{Eq:eta-eng-small-Pgen}).


\subsection{Refrigerator with relaxation}
Our objective is to find an over-estimate of the maximal efficiency of a refrigerator that is made of quantum systems in which relaxation occurs.
The efficiency of the system with relaxation, 
$ \eta_{\rm fri}^{\rm total}(P_{\rm gen})$, is given by 
\begin{eqnarray}
\eta_{\rm fri}^{\rm total}(J_L)  = 
p_{\rm rel} \eta_{\rm fri}^{(1\&2)} (p_{\rm rel}J_L)  
+q_{\rm rel} \eta_{\rm fri}^{(3)} ( q_{\rm rel}J_L), 
\label{Eq:fridges-in-parallel}
\end{eqnarray}
thus we need to find an upper bound on $\eta_{\rm fri}^{(1\&2)}$. 
We make an over-estimate of this efficiency by taking
$T_M$ to be a free parameter between $T_L$ and $T_R$.
For given $T_M$, the efficiency of the combined systems 1 and 2
is
\begin{eqnarray}  
\eta_{\rm fri}^{(1\&2)} (J)= J \Big/ \big[ P_{\rm abs}^{(1)}(J_1) 
+ P_{\rm abs}^{(2)}(J_2) \big],  
\end{eqnarray}
where $J_1=J+J_1^{\rm ph}+J^{\rm ph}$ and $J_2=J+J_2^{\rm ph} +J^{\rm ph}+P_{\rm abs}^{(1)}$,
see Fig.~\ref{Fig:relax}d.
This efficiency is maximized when $J_1^{\rm ph},J_2^{\rm ph},J^{\rm ph} = 0$ 
(since $T_L<T_M<T_R$ means these currents are not negative). Then a little algebra gives 
\begin{eqnarray}
1+{1 \over \eta_{\rm fri}^{(1\&2)}(J)} 
= \left[
1+{1 \over \eta_{\rm fri}^{(1)}(J)} 
\right]\left[
1+{1 \over \eta_{\rm fri}^{(2)}\big(J_2\big)} 
\right] \! , \qquad
\end{eqnarray}
where $J_2= J+P_{\rm abs}^{(1)}= J\big[1+1/\eta_{\rm fri}^{(1)}(J)\big]$.
Thus, to maximize $\eta_{\rm fri}^{(1\&2)}(J)$ for given $T_M$, one must maximize both
$\eta_{\rm fri}^{(1)}$ and $\eta_{\rm fri}^{(2)}$.  
For low power, this can be done using Eq.~(\ref{Eq:eta-fri-smallJ}) 
(much as for the heat-engine in Section~\ref{Sect:eng-eff-relax} above)
giving
\begin{eqnarray}
 \eta_{\rm fri}^{(1\&2)}
 \leq  \eta_{\rm fri}^{\rm Carnot} 
\! \left(\! 1- 1.09
\sqrt{
{T_R \over T_R-T_L}{J_L K_{\rm rel}\over J_L^{\rm qb}(N=1)}}\,\right) \! ,  \  \ 
\end{eqnarray}
where $K_{\rm rel}$ is given in Eq.~(\ref{Eq:Krelax}), and $J_L^{\rm qb}(N=1)$ is given by
Eq.~(\ref{Eq:J-qb-fri}) with $N=1$.
The over-estimate of $\eta_{\rm fri}^{(1\&2)}$ is maximal when $K_{\rm rel}$ is minimal,
see Eq.~(\ref{Eq:Krelax-limit}).
Substituting this into Eq.~(\ref{Eq:fridges-in-parallel}), we see that the efficiency
with relaxation does not exceed the result in Eq.~(\ref{Eq:eta-fri-smallJ})
for a  {\it relaxation-free} system with $N_{\rm max}$ transverse modes.


\subsection{Quantum bounds on power with relaxation}
For a heat-engine,  the arm with scatterers 1 and 2,
has a maximum power, 
\begin{eqnarray}
P_{\rm gen}^{(1\&2)} \leq A_0\, {\pi^2 \over h} \kB^2 
\left[ N_1\big(T_L-T_M\big)^2+ N_2\big(T_M-T_R\big)^2 \right],
\nonumber
\end{eqnarray} 
Since $(T_L-T_M)^2 +(T_M-T_R)^2 \leq (T_L-T_R)^2$, the power of the full system 
cannot exceed the maximum power of a relaxation-less system, Eq.~(\ref{Eq:P-qb2}),
with $N_{\rm max}$ modes.

For a refrigerator, the arm containing scatterers 1 and 2
has a maximum cooling power, 
\begin{eqnarray}
J \leq  \left\{  
\begin{array}{l}
\pi^2  N_1 \kB^2 T_L^2 \big/(12 h) \\
\pi^2 N_2 \kB^2 T_M^2 \big/(12 h) -P_{\rm abs}^{(1)} \ ,
\end{array}\right. 
\end{eqnarray}
where $P_{\rm abs}^{(1)}$ is the electrical power absorbed by scatter 1.
The upper (lower) term is the limit on the heat-flow into scatterer 1 (scatterer 2), 
noting that the heat-flow into scatterer 2 is $J+P_{\rm abs}^{(1)}$.  
Unless $N_2 \gg N_1$, the lower limit  is the more restrictive one.
In any case, the cooling power of the full system 
can never exceed the maximum power of a relaxation-less system, Eq.~(\ref{Eq:J-qb-fri}), with 
$N_{\rm max} $ modes.


\section{Conclusions}
\label{Sect:conclusions}

The upper bound on efficiency at zero power (i.e.~Carnot efficiency) is classical, 
since it is independent of wavelike nature of 
the electrons.  However, this work on thermoelectrics shows that the 
upper bound on efficiency at finite power is quantum,
depending on the ratio of the thermoelectric's cross-section to the electrons' Fermi wavelength.  
If one thought that electrons were classical (strictly zero wavelength), one would believe that Carnot efficiency was achievable at any power output.  Quantum mechanics appears to tell us that this is not so.

However, a crucial point for future work is to discover how universal our bounds on efficiency at 
finite power are.  Our bounds currently rely on the quantum system being (a) well modelled by the 
nonlinear scattering theory with its mean-field treatment of electron-electron interactions,
(b) coupled to only two reservoirs (hot and cold), and (c) relaxation free. 
Under certain conditions we have also shown that they apply when there is relaxation in the quantum system.  
We cannot yet prove that our results are as general as Pendry's bound on heat flow\cite{Pendry1983},
which applies for arbitrary relaxation and for more than two reservoirs \cite{2012w-2ndlaw}, 
as well as for electronic Luttinger liquids\cite{Kane-Fisher} and bosons\cite{Pendry1983}.
It also remains to be seen if our bound occurs in systems with strong electron-electron interactions (Coulomb blockade, Kondo physics, etc.).  
More generally, we wonder whether similar bounds apply to those thermodynamic machines that do not rely on thermoelectric effects, such as Carnot heat engines.

\section{Acknowledgements}

I am very grateful to M.~B\"uttiker for the suggestion which led to the implementation
in Section~\ref{Sect:chain}. I thank P.~H\"anggi for questions on 
entropy flow which led to section~\ref{Sect:Unique}.
I thank  L.~Correa for questions which led to a great improvement of section~\ref{Sect:eff-CA}.  
I thank C.~Grenier for an analytic solution of Eq.~(\ref{Eq:T-for-chain}) for $k=3$.




\begin{thebibliography}{1}
\bibitem{footnote:casati-review}
This point is easily overlooked in plots where power is in units of each system's maximum power,
such as Fig.~4 of Ref.~[\onlinecite{Casati-review}].
It is easy to misread such a plot as implying that the system with $ZT \to \infty$ is the most efficient at all powers.
\end{thebibliography}
\end{document}